\documentclass[12pt]{article}
\usepackage{epsfig}
\usepackage{amsfonts}
\usepackage{amsmath}
\usepackage{amssymb}

\usepackage{color} 
  \newcommand{\beq}{\begin{equation}}
  \newcommand{\eeq}{\end{equation}} 
  \def\nuc#1#2{\relax\ifmmode{}^{#1}{\protect\text{#2}}\else${}^{#1}$#2\fi}
  \def\itnuc#1#2{\setbox\@tempboxa=\hbox{\scriptsize\it #1}
    \def\@tempa{{}^{\box\@tempboxa}\!\protect\text{\it #2}}\relax
    \ifmmode \@tempa \else $\@tempa$\fi}
  
\newcommand{\simge}{\hspace*{0.2em}\raisebox{0.5ex}{$>$}
     \hspace{-0.8em}\raisebox{-0.3em}{$\sim$}\hspace*{0.2em}}
\newcommand{\simle}{\hspace*{0.2em}\raisebox{0.5ex}{$<$}
     \hspace{-0.8em}\raisebox{-0.3em}{$\sim$}\hspace*{0.2em}}

\newcommand{\MQCD}{M_{\mathrm{QCD}}}

\newcommand{\dslash}[1]{#1 \llap{/\kern-0.5pt}}
\newcommand{\Dslash}[1]{#1 \llap{/\kern+1.2pt}}
\newcommand{\DDslash}[1]{#1 \llap{/\kern+2.3pt}}
\newcommand{\dslashh}[1]{#1 \llap{/\kern+1pt}}

\newcommand{\boldtau}{\mbox{\boldmath $\tau$}}

\def\bdm{\begin{displaymath}}
\def\edm{\end{displaymath}}

\begin{document}

\begin{titlepage}

\vspace*{1.5cm}

\begin{center}
{\Large\bf The Problem of Renormalization} 
\\
\vspace{0.3cm}
{\Large\bf of Chiral Nuclear Forces}
\\

\vspace{2.0cm}

{\large \bf U. van Kolck}

\vspace{0.5cm}
{
{\it Universit\'e Paris-Saclay, CNRS/IN2P3, IJCLab,
\\
91405 Orsay, France}
\\
and
\\
{\it Department of Physics, University of Arizona,
\\
Tucson, AZ 85721, USA}
}

\vspace{1cm}

\today

\end{center}

\vspace{1.5cm}

\begin{abstract}
Ever since quantum field theory was first applied
to the derivation of nuclear forces in the mid-20th century,
the renormalization of pion exchange with realistic couplings
has presented a challenge. The implementation of effective field theories (EFTs)
in the 1990s promised a solution to this problem but unexpected 
obstacles were encountered.
The response of the nuclear community has been to focus on ``chiral potentials''
with regulators chosen to produce a good description of data. 
Meanwhile, a successful EFT without explicit pion 
exchange --- Pionless EFT --- has
been formulated where renormalization is achieved order by order in 
a systematic expansion of low-energy nuclear observables.
I describe how lessons from Pionless EFT are being applied to the 
construction of a properly renormalized Chiral EFT.

\end{abstract}

\end{titlepage}

\section{Introduction}
\label{intro}

In the aftermath of the solution of the ``problem of infinities''
in Quantum Electrodynamics (QED), an intense quest set in
to renormalize nuclear forces, where pion exchange replaced 
the photon exchange responsible for atomic forces. 
(For an early example, see Ref. \cite{earlyrenorm}.)
It was quickly understood that the only relativistic
pion-nucleon coupling that is renormalizable is 
pseudoscalar~\cite{Matthews:1951sk}. However, pseudoscalar coupling differs 
from pseudovector coupling by a large nucleon-pair term, which was found
to be in conflict with pion phenomenology~\cite{Marshak:1952NN}.  
For the favored pseudovector coupling, the description of two-nucleon data 
depended sensitively on the high-momentum (or short-distance) cutoff 
(see, for example, Ref. \cite{Gartenhaus:1955zz}). 
Efforts moved towards the investigation of various prescriptions for handling
short-range effects, including specific cocktails of 
(usually single-)heavier-meson exchange, form factors with
{\it ad hoc} shapes, and/or boundary conditions at some finite distance.
Nuclear theory acquired an increasingly phenomenological 
character.
Typically, the nonrelativistic Schr\"odinger equation
was solved with a two-nucleon ($2N$) potential including
one-pion exchange, some approximation to two-pion exchange,
and a more or less arbitrary short-range
form, with sufficiently many parameters to fit data to the desired accuracy.
The end result was that potentials including quite different physics could 
produce very good parametrizations of $2N$ data up to around the 
pion-production threshold, while typically
underpredicting three- and more-nucleon binding by more than 10\%.
A serious difficulty was to infer a satisfactory form of three-nucleon 
($3N$) forces and,
for reactions, $2N$ currents.
Reference \cite{Machleidt:2017vls} recounts some of this history.

In contrast, by the mid-1970s
renormalizable quantum field theories had won
the day in particle physics, leading to the formulation of 
Quantum Chromodynamics (QCD) as the theory
of strong interactions.
Out of the attempts to make predictions for QCD at low energies
and to understand how the Standard Model (SM) can arise from
a more fundamental theory, the concept of 
effective field theory (EFT) was born \cite{Weinberg:1996kw}. 
An EFT comprises all the interactions among relevant degrees of
freedom that are allowed by symmetries, including
an arbitrary number of fields and derivatives.
For predictions, contributions to observables must be ordered
according to their expected size.
This ``power counting'' allows for an {\it a priori} error estimate
from neglected higher-order contributions.
At each order in the expansion, only a finite number of 
``low-energy constants'' (LECs) --- the interaction strengths --- appear. 
In a consistent power counting, they are sufficient to
ensure that any dependence on the regulator can be made arbitrarily 
small by taking the cutoff large. 
Thus, EFTs are renormalizable in the modern sense that 
at each order a finite number of parameters generate results for observables 
that are independent 
of details of the arbitrary regularization procedure.

A successful EFT, Chiral Perturbation Theory (ChPT), was developed in the 1980s
to handle interactions among pions and one nucleon below
the characteristic QCD scale $M_{\rm QCD} \sim 1$ GeV 
\cite{Weinberg:1978kz,Gasser:1983yg}.
Requiring renormalization in a perturbative expansion, 
a consistent power counting was developed 
based on ``naive dimensional analysis'' (NDA) \cite{Manohar:1983md}.
Taking the typical external momentum in a reaction to be 
of the order of the pion mass,
$Q\sim m_\pi \ll M_{\rm QCD}$, observables are expanded in a series of
powers of $Q/M_{\rm QCD}$ times calculable functions of $Q/m_\pi$.
When Weinberg remarked \cite{Weinberg:1990rz,Weinberg:1991um} that ChPT, now 
generalized as ``Chiral EFT'' (ChEFT), could be used to derive nuclear forces,
he identified an infrared enhancement in nuclear amplitudes
by the nucleon mass $m_N={\cal O}(M_{\rm QCD})$,
which can lead to 
the failure of perturbation theory --- a good thing since nuclei are
bound states and resonances. 
He proposed that the ChPT power counting
could still be applied to the nuclear potential, defined as 
the sum of diagrams lacking an explicit enhancement.
Then, the Lippmann-Schwinger equation, or equivalently
the Schr\"odinger equation, would be 
solved with a truncated ``chiral potential''.

The potential defined by Weinberg contains pion exchange diagrams
where all LECs are fixed, at least in principle, from ChPT.
But it also includes shorter-range interactions with LECs that can
only be determined in nuclear systems.
Implicit in Weinberg's proposal was that the short-range LECs
would not contain
an {\it implicit} enhancement. This would be the case if the solution
of the dynamical equation does not generate
cutoff dependence beyond that which can be compensated
by the LECs already present up to that order according to NDA.

Whether this assumption is true was not immediately clear. NDA 
says that the potential at leading order (LO) consists of
two non-derivative, chirally symmetric 
contact interactions together with one-pion exchange (OPE).
More-pion exchange should come at higher orders together with more-derivative
contact interactions.
Nonperturbative pion exchange prevents an analytical solution even at the 
$2N$ level.
The first numerical solution of a chiral potential in the 
$2N$ system \cite{Ordonez:1993tn,Ordonez:1995rz}
tested renormalizability of the amplitude:
a variation from 0.5 to 1 GeV in the cutoff of a local Gaussian 
regulator 
seemed to be compensated by a refitting of the
LECs at hand.
However, the fitting procedure was cumbersome as 
an over-complete set of interactions
was used and the local regulator
mixed different partial waves, limiting the range of cutoffs 
that could be explored.
Since then a large variety of chiral potentials have been developed 
(for reviews, see for example Refs. \cite{Epelbaum:2008ga,Machleidt:2011zz}).
A landmark was a $2N$ potential \cite{Entem:2003ft} that was 
perceived to match the accuracy of phenomenological potentials
(for a recent comparison between chiral $2N$ potentials 
and data, see Ref. \cite{NavarroPerez:2019sfj}).
Chiral potentials have become the favorite input to ``{\it ab initio}'' 
methods, which provide numerically controlled solutions of the
Schr\"odinger equation for multi-nucleon systems.

Unfortunately, pretty early on the first evidence appeared \cite{Kaplan:1996xu}
that Weinberg's prescription does not provide amplitudes, and thus observables,
that are renormalized order by order. 
In the $2N$ $^1S_0$ channel at LO, a semi-analytical argument
shows that there remains a logarithmic dependence on the cutoff proportional 
to the average quark mass. The only way to eliminate it, at least with
a momentum- or coordinate-space cutoff, is to include at LO
a non-derivative, chirally {\it breaking}
contact interaction, which according to NDA should appear two orders
down the expansion, that is, at next-to-next-to-leading order (N$^2$LO) 
\footnote{A note on notation:
It has become usual in the nuclear community to refer to a 
subleading chiral potential of order $n\ge 2$ 
as ``N$^{n-1}$LO'',
because with Weinberg's power counting the parity- and time-reversal-invariant
potential of order $n=1$ vanishes \cite{Ordonez:1992xp}. 
However, this usage is too provincial
to accommodate experience with other observables and power countings
in ChEFT or other EFTs. Here,
a correction of order $n$ in the expansion is denoted as N$^{n}$LO,
whether it is non-zero or not.}.
More dramatically, it was later shown \cite{Nogga:2005hy,PavonValderrama:2005uj}
that oscillatory
cutoff dependence appears at LO in waves where OPE is 
attractive, singular, and accounted for nonperturbatively.
A chirally symmetric LEC is needed for renormalization in each wave,
but again NDA assigns those in partial waves beyond $S$ to higher orders.
Similar problems afflict processes with external probes 
\cite{Valderrama:2014vra}.

As I describe in Sec. \ref{sing}, the origin of these problems
is the renormalization of attractive singular potentials 
\cite{Beane:2000wh,PavonValderrama:2007nu}.
NDA might fail because exact solutions of the Schr\"odinger equation 
depend on the cutoff differently than perturbative solutions.
The LECs needed for the renormalization of the amplitude {\it are} 
enhanced by implicit powers of $M_{\rm QCD}$.

How to account for this? In response to the renormalization failure
of Weinberg's power counting a simpler nuclear EFT 
\cite{Bedaque:1997qi,vanKolck:1997ut,vanKolck:1998bw}
was developed in the late 1990s. 
In this ``Pionless EFT'' pions are integrated out and only
contact interactions remain. The effects of loops in the Lippmann-Schwinger
equation are much easier to see, including the $m_N$ enhancement
and a further enhancement of $4\pi$ \cite{vanKolck:1997ut,vanKolck:1998bw}
that was not pointed out by Weinberg. 
The lessons of Pionless EFT for ChEFT are summarized in Sec. \ref{lessons}.

The first attempt to fix power counting using 
the insights from Pionless EFT
was initiated \cite{Kaplan:1998tg,Kaplan:1998we} at the same
time as the main elements of the power counting of Pionless EFT were 
being understood.
Valid for sufficiently small values of the pion mass and external momenta,
this version of ChEFT treats pion exchange in perturbation theory,
removing the renormalization problems mentioned above.
Unfortunately, in the 
$2N$ system at physical pion mass one cannot go in this way to momenta
much beyond those of Pionless EFT 
\cite{Fleming:1999ee}. 
The alternative is {\it partly} perturbative pions: OPE is iterated only
in the low partial waves where it is sufficiently strong, together
with the contact interactions whose LECs are necessary for LO renormalization
\cite{Nogga:2005hy}.
All subleading pion exchanges, together with the remaining contact interactions,
are treated in perturbation theory \cite{Long:2007vp}. 
This approach is discussed in Sec. \ref{RenormChi},
including what little has been done to confront it with data.

Section \ref{conc} offers the conclusion that
this approach solves the renormalization
woes of nuclear forces while accounting for the long-range interactions
from pion exchange systematically.
Although they differ in detail from the 
field-theoretical renormalization
described below, renormalization-group analyses of the Schr\"odinger 
equation 
\cite{Birse:2005um,Birse:2010fj,Valderrama:2014vra,Valderrama:2016koj}
support this picture.
How it can meet the accuracy requirements of the nuclear community
remains to be seen.
My emphasis here is on the {\it internal consistency} of ChEFT.
I expand on the renormalization issues summarized in Ref. \cite{Hammer:2019poc},
but I refer the reader to the latter for a more complete review of ChEFT 
and its relation to other nuclear EFTs.

\section{Say what?}
\label{lessons}

As reviewed in Ref. \cite{Hammer:2019poc},
defining the nuclear potential as the sum of ``irreducible'' diagrams
without the $m_N$ infrared (IR) enhancement
does indeed ensure that the {\it cutoff-independent} parts of
pion-exchange diagrams can be ordered according to ChPT power counting.
These components of the pion-exchange potentials
are in general non-analytic functions of momenta and pion mass
that can be calculated in terms of pion-baryon interactions.

The ChPT power counting is designed for processes where the typical external 
momentum is comparable to the pion mass, $Q\sim m_\pi$.
A (relativistic) pion propagator scales as $Q^{-2}$. 
In contrast, a nucleon is heavy compared to $Q$ and thus
nonrelativistic. Moreover, energies
and three-momenta being comparable,
nucleon recoil is suppressed by one power of $Q/m_N={\cal O} (Q/\MQCD)$
--- that is, the nucleon is static,
its propagator scaling as $Q^{-1}$.
Because the Delta-nucleon mass difference is (at physical quark masses)
only about twice the pion mass, a Delta propagator scales in the same way.
In integrals from the loops that make up the potential one picks poles from the 
pion propagators, typically resulting in factors of $(4\pi)^{-2}$. 
They combine with factors of the pion decay constant $f_\pi\simeq 92$ MeV
from the pion-baryon interactions 
to produce inverse factors of
$4\pi f_\pi ={\cal O}(M_{\rm QCD})$.
The power counting explicitly relies on an estimate, NDA \cite{Manohar:1983md},
of the factors of $4\pi$
that distinguish between $f_\pi$ and the breakdown scale $M_{\rm QCD}$,
which appears in interactions with derivatives and powers of 
the pion mass. 
In summary, the ChPT rules (in momentum space) are:
\begin{eqnarray}
 \text{(pion) loop integral} &\sim& (4\pi)^{-2}Q^4 \,,
\label{loopintegral}\\
 \text{baryon, pion propagator} &\sim& Q^{-1}, Q^{-2} \,,
\label{props} \\
 \text{vertex} &\sim& Q^d f_\pi^{2-b-f} \MQCD^{2-d-f/2}\,,
\label{vertices}
\end{eqnarray}
where $d$, $b$, and $f$ are the numbers of derivatives/pion masses, 
pion fields, and baryon fields, respectively, in an interaction.

The expected size of any diagram can be found
using the identities $I=L-1+\sum_i V_i$ and $2I+E=\sum_i V_i (b_i+f_i)$ involving
the number of loops ($L$), internal (external) lines $I$ ($E$),
and vertices ($V_i$) having a set of values $d=d_i$, $b=b_i$, and $f=f_i$.
In particular,
\begin{equation}
\text{$2N$ potential}
\sim 4\pi m_N^{-1} M_{N\!N}^{-1}\left(Q \MQCD^{-1}\right)^{\mu} \,,
\label{2Npotsize}
\end{equation}
where \cite{Kaplan:1998tg,Kaplan:1998we}
\begin{equation}
M_{N\!N} \equiv \frac{16\pi f_\pi^2}{g_A^2 m_N} = {\cal O}(f_\pi)
\label{OPEscale}
\end{equation}
in terms of the pion-nucleon axial-vector coupling $g_A\simeq 1.27$
and  \cite{Weinberg:1991um}
\begin{equation}
\mu \equiv 2L + \sum_i V_i (d_i+f_i/2 -2) \,.
\label{Wpc}
\end{equation}
Because every additional loop (without increase in the number of 
derivatives/pion masses at vertices) leads to a 
relative factor ${\cal O}(Q^2/M_{\rm QCD}^2)$, 
one gets the well-known ordering where 
$p$-pion exchange starts at $\mu=2(p-1)$.
Note that the NLO correction vanishes due to parity and time-reversal
symmetries \cite{Ordonez:1992xp}.

\begin{figure}[t]
\begin{center}
\includegraphics[scale=.6]{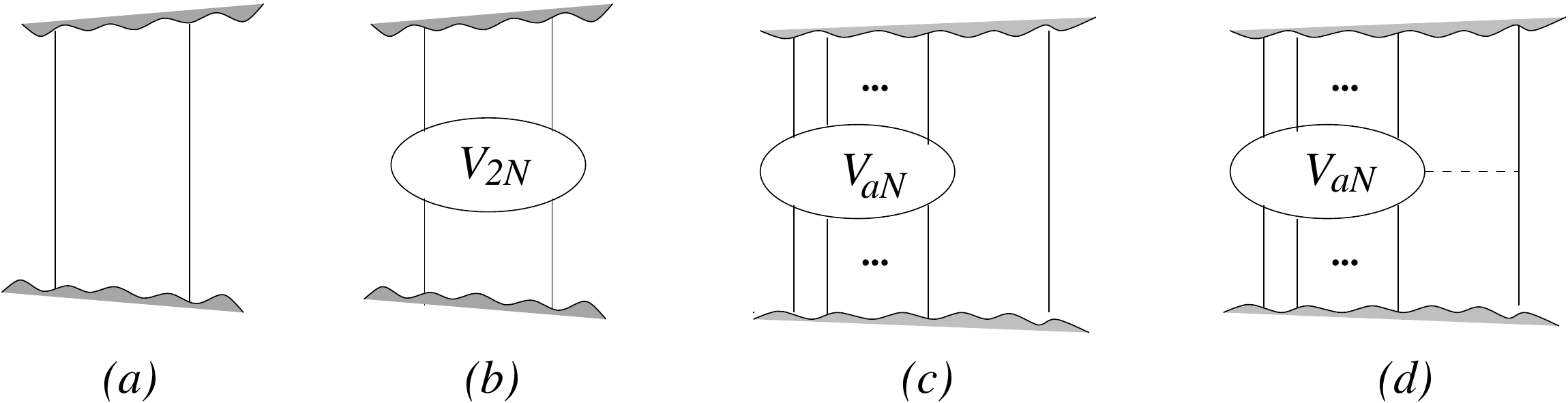}
\end{center}
\caption{
Some diagrams discussed in the text.
Inside a diagram,
{\it (a)} two nucleons (solid lines) propagate;
{\it (b)} two nucleons interact through the $2N$ potential (blob);
{\it (c)} $a$ nucleons interact through the $aN$ potential (blob),
while another nucleon propagates;
and
{\it (d)} $a+1$ nucleons interact through the $(a+1)N$ potential
formed from the $aN$ potential and the exchange of a pion (dashed line).
}
\label{somediagrams}
\end{figure}

This power counting applies to diagrams that make up the long-range potential.
Yet physics, as opposed to metaphysics, is about observables.
The meaning of Eq. \eqref{2Npotsize} is
that it indirectly orders the contributions to {\it amplitudes}.
For the direct link, we need to consider as well ``reducible'' diagrams
where intermediate states contain only nucleons.
One picks poles from the nonrelativistic nucleon propagators,
for which energies are of the order of recoil --- in those diagrams,
one can{\it not} approximate nucleons as static.
(This of course has nothing to do with relativistic corrections,
as sometimes misstated in the literature.)
These poles lead not only to 
an $m_N$ enhancement \cite{Weinberg:1990rz,Weinberg:1991um}, 
but typically also to different powers of $(4\pi)^{-1}$.
Experience with Pionless EFT
\cite{Bedaque:2002mn,Hammer:2019poc}, 
where these are all the loops one needs to deal with,
shows that the factors associated with reducible
loops are
\begin{eqnarray}
\text{nucleon propagator} &\sim& m_N Q^{-2} \,, 
\label{red1} \\
\text{reducible loop integral} &\sim& (4\pi m_N)^{-1} Q^5 \,.
\label{red2} 
\end{eqnarray}
When one inserts the order-$\mu$ potential into a $2N$ diagram
we need one extra reducible loop with two nucleon propagators
(compare Figs. \ref{somediagrams}{\it (a)} and {\it (b)}),
leading to a relative factor $(Q/M_{N\!N})(Q/M_{\rm QCD})^{\mu}$.  
This amount to an IR enhancement of $4\pi m_N/Q$ over the factor that 
arises from Eqs.  \eqref{loopintegral} and \eqref{props}.
As a consequence, the series in
the LO potential fails to converge for $Q\sim M_{N\!N}$.
This is what makes ChEFT different for $A\ge 2$ nucleons
compared to ChPT for $A\le 1$. 

The factor of $4\pi$ in the IR enhancement had not been
recognized before Pionless EFT was developed,
but it is important to understand the failure of perturbation
theory {\it for pions}.
The exact solution of the LO potential for $Q\sim M_{N\!N}$ can give rise
to a binding energy per nucleon
\begin{equation}
\frac{B_A}{A}\sim \frac{M_{N\!N}^2}{\MQCD}
\sim \frac{f_\pi}{4\pi} \sim 10 \; {\rm MeV} \, .
\label{BA}
\end{equation}
This is somewhat larger than observed for light nuclei,
indicating a certain amount of fine tuning in the $2N$ interactions.
But it is on the right ballpark for heavier nuclei, so 
chiral symmetry together
with the IR enhancement provides a natural explanation~\cite{Bedaque:2002mn}
for the shallowness of nuclei compared to $\MQCD$, $B_A/A\ll \MQCD$, long
considered a mystery.

The same factor of $4\pi$ has implications for the natural size of few-body 
forces, which were recognized by Friar \cite{Friar:1996zw}. 
To see this, consider connecting a nucleon with OPE
to an $aN$ potential to make an $(a+1)N$ potential,
without changing the number of derivatives, pion masses,
and loops in the $aN$ potential.
(See Figs. \ref{somediagrams}{\it (c)} and {\it (d)}.
For example, take the crossed-box two-pion exchange $2N$ potential 
and connect one of the intermediate nucleons to the third nucleon.)
The additional nucleon propagator inside the $aN$ potential and
the additional OPE combine for a factor
of $4\pi m_N^{-1} M_{N\!N}^{-1}Q^{-1}$.
At the same time, at the amplitude level we are adding a reducible 
loop and one propagator for the extra nucleon, that is, 
another factor $(4\pi)^{-1}Q^3$.
The contribution of the $(a+1)N$ potential to the amplitude
is, overall, of ${\cal O}(Q^2 m_N^{-1} M_{N\!N}^{-1})$ compared to that
of its ``parent'' $aN$ potential.
For $Q\sim M_{N\!N}$, the suppression from connecting a nucleon is 
thus of ${\cal O}(Q/M_{\rm QCD})$,
or one order in the expansion of the potential \cite{Friar:1996zw}.  
In contrast, missing the $4\pi$ in the IR enhancement
would give an additional $(4\pi)^{-1}={\cal O}(M_{N\!N}/M_{\rm QCD})$,
or a suppression of $(Q/M_{\rm QCD})^2$ 
\cite{Weinberg:1991um,Ordonez:1992xp,Weinberg:1992yk,vanKolck:1994yi}.
In either case a hierarchy of many-body forces arises,
with perturbative $3N$ forces coming after the leading $2N$ forces.
Unfortunately, existing calculations do not question
the additional suppression of $(4\pi)^{-1}$.

\begin{figure}[t]
\begin{center}
\includegraphics[scale=.7]{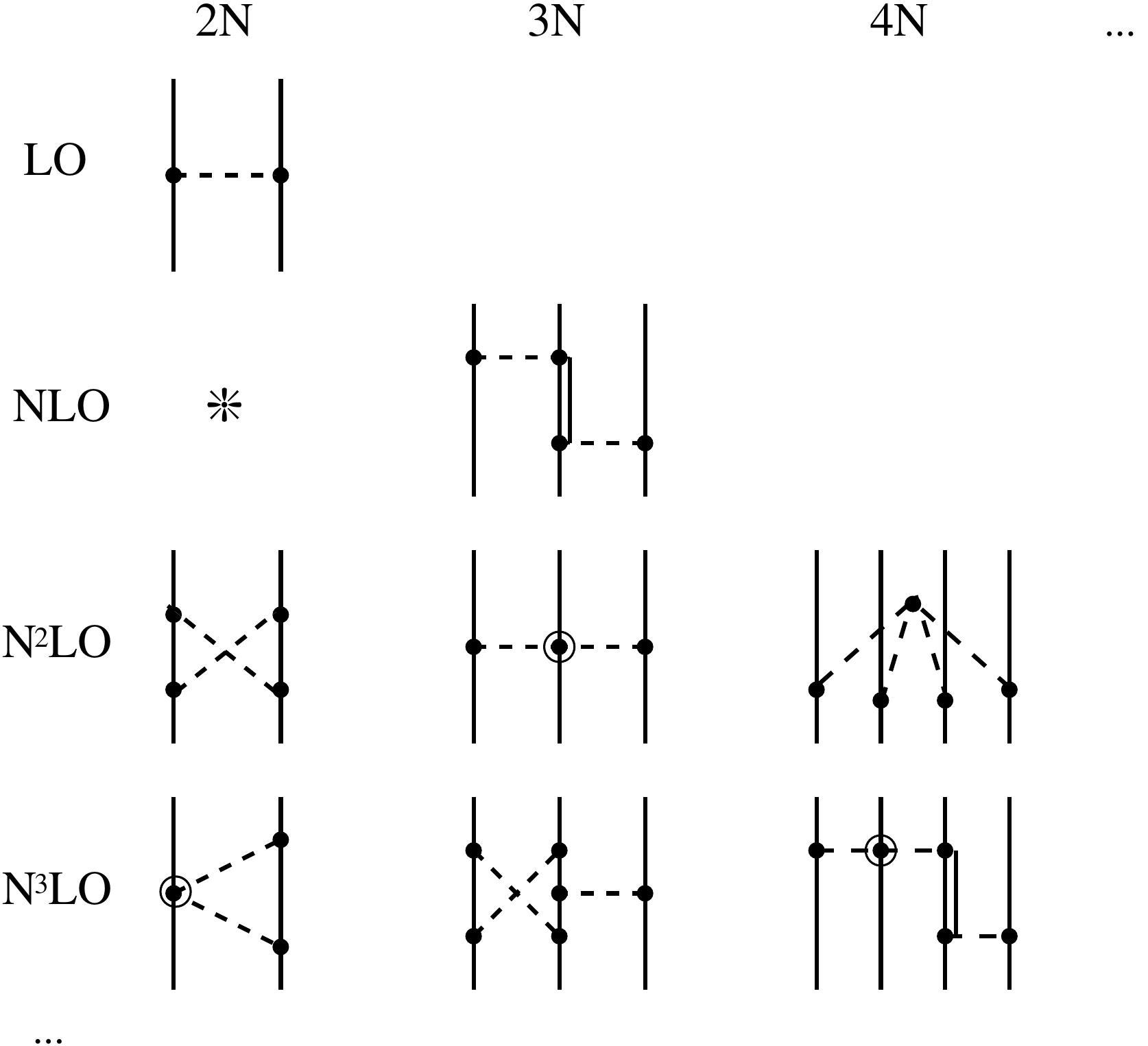}
\end{center}
\caption{Sample of pion-range diagrams in
the $aN$ nuclear potential ordered according to the expected size
of their contributions to the amplitude, Eq. \eqref{Fpc}.
N$^{\nu}$LO denotes relative ${\cal O}(Q^{\nu}/\MQCD^{\nu})$.  
A solid (double) line stands for a nucleon (nucleon 
excitation), while a dashed line, for a pion.  A circle (circled circle)
represents an interaction with $d+f/2-2=0$ ($=1$).
}
\label{chiralpot}
\end{figure}

Note that when connecting the additional nucleon we might not be able
to maintain the number of derivatives or pion masses.
In particular, for the leading $aN$ force, this can only be done with
an intermediate Delta isobar --- for $3N$, that is the Fujita-Myiazawa
force \cite{Fujita:1957zz},
which has been argued to be important for convergence of the chiral
expansion \cite{Pandharipande:2005sx}.
Keeping this in mind, a contribution to the (connected) $aN$ potential scales as
\begin{equation}
\text{$aN$ potential}
\sim (4\pi m_N^{-1} M_{N\!N}^{-1})^{a-1} Q^{2-a} \left(Q \MQCD^{-1}\right)^{\mu} \,.
\label{aNpotsize}
\end{equation}
To estimate the respective contributions to the $AN$ amplitude,
one can first consider the LO ($\mu=0$), $2N$ potential:
to produce a connected diagram, we need at least $A-1$ $2N$ interactions
linked by $A-2$ propagators. 
Next, one insertion of a subleading $aN$ potential between two LO amplitudes 
comes with $A+a-2$ propagators and $A+a$ loops.
Another insertion of the same subleading potential 
takes $a$ additional propagators and $a-1$ additional loops.
And so on.
The rules \eqref{red1}, \eqref{red2} imply that 
an $aN$ potential
of index $\mu$ gives, at $Q\sim M_{N\!N}$,
\begin{equation}
\text{$AN$ amplitude} 
\sim (4\pi)^{A-1} m_N^{-1} M_{N\!N}^{5-3A} \left(Q \MQCD^{-1}\right)^{n\nu} \,,
\label{subLOANampsize}
\end{equation}
where 
\begin{equation}
\nu \equiv \mu + a -2
\label{Fpc}
\end{equation}
and $n$ is the order in perturbation theory.
While $\nu$ is the perturbative cost of one insertion of 
a subleading potential characterized by $\mu$ \eqref{Wpc} and $a$,
$n$ insertions cost $n\nu$ as indicated by the power of $Q/\MQCD$
in Eq. \eqref{subLOANampsize}.
The presence of $a-2$ (instead of $2(a-2)$) in $\nu$ reflects the suppression
by $(4\pi)^{-1}$ (instead of $(4\pi)^{-2}$) in more-nucleon forces.
A sample of pion-range diagrams that contributes at various
values of $\nu$ is shown in Fig. \ref{chiralpot}, 
see Ref. \cite{Hammer:2019poc} for more details.

The $n$ in Eq. \eqref{subLOANampsize} encodes the perturbative 
character of any subleading interaction. 
A common fallacy is that the mere definition of a potential means that the 
corresponding 
dynamical (Lippmann-Schwinger or Schr\"odinger) equation must be solved exactly.
On the contrary, if there is a sense in which a subleading potential can
be treated nonperturbatively, then it should also be possible to include
it in distorted-wave perturbation theory, where the distortion
is caused by the LO potential.
If that is not the case, then at least part of that ``subleading''
potential is not subleading.
Such a consistency test is almost completely ignored in the community.
The one exception I am aware of is Ref. \cite{Valderrama:2019lhj},
where it is shown that this test
is not met by most available chiral potentials.

``But surely'', you might be reasoning, ``a subleading potential 
{\it can} be treated nonperturbatively.'' 
That is certainly the case for a regular subleading potential, 
but not necessarily
for a singular potential, for which neither the perturbative series nor
the exact solution of the dynamical equation are well defined
without (potentially distinct) counterterms.
So far I have been glossing over the cutoff dependence that usually
arises in loops and is, of course, present in the LECs.
A regulator is nothing but a way to split 
short-range physics between loops and LECs. 
If we increase a momentum cutoff $\Lambda$
(or decrease a coordinate cutoff $R\sim \Lambda^{-1}$),
we account, correctly or incorrectly,
for more short-range physics through the loops
of the Lippmann-Schwinger equation.
As long as $\Lambda\simge M_{\rm QCD}$, we can compensate by
changing the LECs present at the same order, without increasing
the relative truncation error of ${\cal O}(Q/M_{\rm QCD})$.
The crucial point is that only the combination of the two effects matter, 
and physics enters through the fitting of as many observables as LECs 
--- observables which are either calculated in the underlying theory 
(when we speak of ``matching'' the EFT to the underlying theory) 
or measured experimentally.
This process of renormalization is essential for
{\it amplitudes}
to be free of detailed assumptions about short-range physics, and 
in general only the sum of all contributions at a given order 
--- loops and LECs ensuring renormalization --- can be said 
to be perturbative or not.

If all we needed was to eliminate the cutoff-dependent parts of 
pion exchange in the potential,
the LECs for the job would be given by NDA, 
by construction \cite{Manohar:1983md}. 
It is crucial to realize, though, that reducible loops
introduce further cutoff dependence, which we need eliminate as well.
The potential itself
has to depend on the cutoff so that observables do not.
The LECs that renormalize this part of the $A\ge 2$ problem
will not in general satisfy NDA.
We examine this aspect of renormalization next.

\section{Renormalization of singular potentials}
\label{sing}

The difficulty we face is that EFT potentials are singular and, because
of additional derivatives and loops, they get
more and more singular as the order of the EFT expansion increases. 
Singularities are apparent already in 
the LO ($\mu=0$, $a=2$) pion-range potential, OPE:
labeling the two nucleons 1 and 2,
\begin{equation}
V_{\rm OPE}(\vec{r}) = \frac{\boldtau_1\cdot \boldtau_2 }{m_N M_{N\!N}} \,
\left[ 
\frac{e^{-m_\pi r}}{r^3} \left(1+m_\pi r+\frac{m_\pi^2 r^2}{3}\right) S_{12}(\hat{r}) 
+ \left(m_\pi^2\frac{e^{-m_\pi r}}{r} -4\pi \delta(\vec{r})\right)
\frac{\vec{\sigma}_1\cdot \vec{\sigma}_2}{3}\right]
\, , 
\label{OPE}
\end{equation}
where $\vec{r}=r \hat{r}$ is the relative position,
$\vec{\sigma}_i$ ($\boldtau_i$) is the spin (isospin) Pauli matrix
for nucleon $i$, and
\begin{equation}
S_{12}(\hat{r})=3\,\vec{\sigma}_1\cdot\hat{r}\,\vec{\sigma}_2\cdot\hat{r}
-\vec{\sigma}_1\cdot \vec{\sigma}_2 
\end{equation}
is the spin-tensor operator.
While the delta function contributes only to $S$ waves,
the tensor potential is non-vanishing for total spin $s=1$ and can mix
waves with orbital angular momentum $l=j\pm 1$.  
It is attractive in some uncoupled waves like $^3P_0$ 
and $^3D_2$, and in one of the eigenchannels of each coupled wave.  
The regular Yukawa 
potential is attractive in isovector (isoscalar) channels for $s=0$ ($s=1$).
More-pion exchange leads to more singular terms,
$p$-pion exchange containing for example terms $\propto r^{-(2p+1)}$
in addition to delta functions and their derivatives.

For $Q\sim M_{N\!N}$ OPE is expected to be nonperturbative
by the argument of the previous section.
It has been known
for a long time (see, {\it e.g.}, the review \cite{Frank:1971xx}) 
that attractive singular potentials, treated exactly,
do not fully determine the solution of the Schr\"odinger equation
\cite{Case:1950an}.
This is a manifestation that renormalization of a singular potential
requires contact terms that naturally exist in EFT 
\cite{Beane:2000wh,PavonValderrama:2007nu}.
In contrast, pion-range corrections to OPE are expected to be perturbative
according to 
the power counting embodied in Eqs. \eqref{subLOANampsize} and \eqref{Fpc}.
From an EFT perspective,
additional contact interactions are needed to make these corrections
well defined \cite{Long:2007vp}.

The issue I address in this section is how many, and which, contact
interactions must be present for the renormalization
of specific singular potentials.
For simplicity, I consider central potentials; we return to
the nuclear potential in Sec. \ref{RenormChi}.

\subsection{Nonperturbative renormalization}
\label{singnonpert}

Renormalization is usually discussed at the level of loops in Feynman
diagrams for the Lippmann-Schwinger equation in momentum space,
but it can also be formulated in terms of the Schr\"odinger equation
in coordinate space. 
In the latter, which is more familiar to many,
renormalization deals with distances on the order of those
where the EFT breaks down, which I will call $R_{\rm und}$. 
The fall off of the 
potential at much larger distances is not important,
as it affects instead the near-threshold behavior.
For definiteness, let us take a central two-body potential 
\begin{equation}
V_{\rm L}(r)= -
\frac{\alpha}{2\mu r^n} 
\label{singpot}
\end{equation}
in the center-of-mass frame, where $\mu$ is the reduced mass, 
$\alpha$ is a constant with mass dimension $2-n$, 
and $n>0$ is an integer.
The long-range potential is characterized by an intrinsic distance
scale $r_0\equiv |\alpha|^{1/(n-2)}$.
For $n=2$ the action is scale invariant.

In the radial Schr\"odinger equation the potential is supplemented
by the centrifugal barrier with orbital angular momentum $l$,
$l(l+1)/(2\mu r^2)$.
The uncertainty principle implies the kinetic
term scales similarly, as $1/(2\mu r^2)$. 
For $0<n<2$ the potential is relatively small at small distances
and the corresponding behavior of the wavefunction is determined
by $l$: we find ourselves in the familiar situation where 
one solution, labeled regular, behaves as $r^l$ for small $r$,
while the other, labeled irregular and discarded, as $r^{-(l+1)}$.
In contrast, for $n=2$ and $|\alpha|$ is sufficiently large, or for $n\ge 3$, 
$V_{\rm L}(r)$ dominates at small distances.
If $\alpha<0$,
the strong repulsion prevents any short-range approach;
one can again keep just the regular solution, from which the scattering
amplitude can be calculated.
But when the potential is attractive, $\alpha>0$,
observables are sensitive to short-distance physics
and renormalization is needed.

To see this in detail, consider first $n\ge 3$ at zero energy.
For $r \simle [l(l+1)]^{-1/(n-2)} r_0$,
where $V_{\rm L}(r)$ dominates,
the Schr\"odinger equation 
becomes an ordinary Bessel equation, and the solution
is a combination of spherical Bessel functions. 
Both solutions
are equally irregular as $r\to 0$ \cite{Case:1950an}.
One can write the wavefunction in the $l$ wave at small distances as
\begin{equation}
\psi_l(r)\propto r^{n/4-1} \cos\left(\frac{\sqrt{\alpha}\, r^{1-n/2}}{n/2-1}
 + \phi_{l}\right) + \ldots \, ,
\label{wfsingpot}
\end{equation}
where $\phi_l$ is a phase that determines the relative importance of the
two irregular solutions and is {\it not} fixed by the long-range potential 
$V_{\rm L}$. This is in strong contrast with the repulsive case, where
the solutions are regular and irregular modified Bessel functions,
which respectively decrease and increase exponentially as $r$ decreases.

The case $n=2$ is borderline singular, 
the character of the solution depending on the relative
size of $\alpha$ and a combination of $l(l+1)$ with a number ${\cal O}(1)$
coming from the kinetic repulsion. 
It turns out that the critical
value is $\alpha_{l}=(l+1/2)^2$.
For $l\ge l_{\alpha}\equiv \sqrt{\alpha}-1/2$, repulsion wins;
one solution is more singular than the other and can again be discarded
\cite{landau}. 
For $l<l_{\alpha}$ the situation is similar to 
$n\ge 3$: Eq. \eqref{wfsingpot} holds with 
$\sqrt{\alpha} \, r^{1-n/2}/(n/2-1)\to 
\sqrt{\alpha -\alpha_{l}} \,\ln(r/r_0)$, 
where $r_0$ is an arbitrary dimensionful parameter and
$\phi_l=\phi_l(r_0)$.  
This is an example of an
anomaly~\cite{Camblong:2003mz,Camblong:2003mb} 
where the scale invariance of the 
classical system 
is broken by the renormalization of the quantum system.

Equation \eqref{wfsingpot} is the quantum version of the ``fall to the center''
in a classical singular potential \cite{landau,Perelomov:1970fz}.
The phases $\phi_l$ determine the asymptotic behavior of the wavefunction,
from which the zero-energy scattering amplitude is extracted.
For example, the $S$-wave scattering length is well defined for
a pure $n\ge 4$ potential \cite{Perelomov:1970fz} and given for $n=4$ by
\begin{equation}
a_0=\sqrt{\alpha} \, \tan \phi_0 \, .
\end{equation}
If one imposes a particular value on $\psi_l(R)$ at a chosen distance $R$
--- for example, that the wavefunction $\psi_l(R)=0$ ---
the phases are fixed.
However, a different value of $R$ 
leads to different phases.
In EFT, this arbitrariness is replaced by the values of LECs.
The minimal set of contact interactions is determined by 
demanding renormalizability.

\subsubsection{$S$ wave}
\label{singnonpertS}

Let us look into the $S$ wave first.
Choosing a sharp cutoff in coordinate space at $R$, we replace 
the potential \eqref{singpot} by \cite{Beane:2000wh}
\begin{equation}
V(r)= V_{\rm S}(R) \, \theta(R-r) + V_{\rm L}(r) \, \theta(r-R) \, .
\label{singpotcomplete}
\end{equation}
The depth $V_{\rm S}(R)$ of the spherical well is related to the LEC $C_0$ 
of a contact interaction,
\begin{equation}
C_0\,\delta(\vec{r})= \frac{C_0}{4\pi r^2} \,\delta(r)
\to \frac{3C_0(R)}{4\pi R^3} \,\theta(R-r)
\equiv V_{\rm S}(R) \, \theta(R-r) \, .
\label{contact}
\end{equation}
A solution of the Schr\"odinger equation
for the augmented potential requires the matching
of the logarithmic derivatives of outside 
and regular spherical-well wavefunctions at $r=R$,
\begin{equation}
\sqrt{-2\mu R^2 V_{\rm S}(R)} \, \cot \sqrt{-2\mu R^2 V_{\rm S}(R)}
=\left.r\frac{\partial}{\partial r} \ln \left(r \psi_0(r)\right)
\right|_{r=R} \,.
\label{match}
\end{equation}

When $n=2$ and $\alpha\le \alpha_{0}$, or $n=1$,
we can solve this equation with $V_{\rm S}(R)=0$ 
if the admixture of the most singular external 
solution tends to zero as $R\to 0$. 
Thus the amplitude is renormalized properly without a contact interaction
as long as we retain only the least singular wavefunction behavior, 
the prescription offered in Ref. \cite{landau}.

\begin{figure}[t]
\begin{center}
\includegraphics[scale=.9]{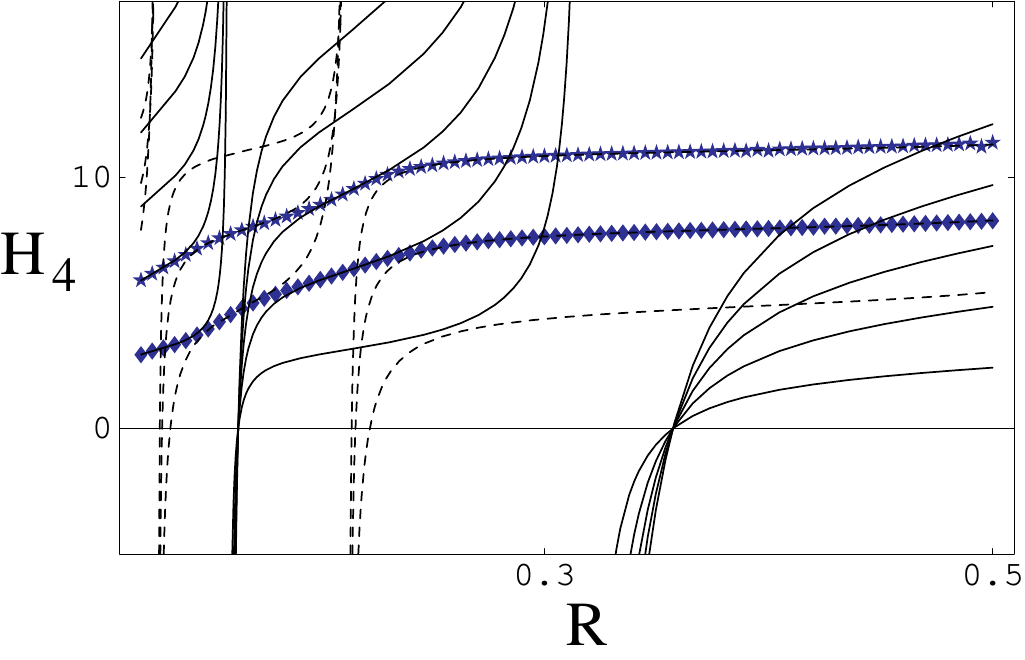}
\end{center}
\caption{Dependence of $H_4\equiv \sqrt{-2\mu R^2 V_{\rm S}(R)}$ for $n=4$
on $R$ (in units of $r_0$).
Two analytical approximations, Eq. \eqref{matchlarge} (solid lines)
and Eq. \eqref{matchsmall} (dashed lines), are shown
together with a numerical solution of Eq. \eqref{match} (bold lines)
that interpolates between them.
Reprinted figure with permission from Ref. \cite{Beane:2000wh}.
Copyright (2001) by the American Physical Society.}
\label{contactforr-2}
\end{figure}

For $n=2$ and $\alpha>\alpha_{0}$,
or for $n\ge 3$,
because the two external solutions differ only by a phase, 
the contact interaction is necessary.
Substituting the wavefunction \eqref{wfsingpot} into Eq. \eqref{match}, 
yields a transcendental equation linking $\phi_0$ to $V_{\rm S}(R)$ 
\cite{Beane:2000wh}. 
Two approximate solutions are
\begin{equation}
\sqrt{-2\mu R^2 V_{\rm S}(R)} \simeq 
m\pi \left\{ 1-\left[1-\frac{n}{4}+
\sqrt{\alpha}\, R^{1-n/2}
\tan\left(\frac{2\sqrt{\alpha}}{n-2}R^{1-n/2}+\phi_0\right)\right]^{-1}\right\}\, ,
\label{matchlarge}
\end{equation}
when the right-hand side of Eq. \eqref{match} is large,
and
\begin{equation}
\sqrt{-2\mu R^2 V_{\rm S}(R)} \simeq 
\frac{(1+2m)\pi}{2} 
-\frac{2}{(1+2m)\pi}\left[\frac{n}{4}-
\sqrt{\alpha}R^{1-n/2}
\tan\left(\frac{2\sqrt{\alpha}}{n-2}R^{1-n/2}+\phi_0\right)\right]\, ,
\label{matchsmall}
\end{equation}
when it is small, where in both cases $m$ is an integer.
Now one can keep the scattering 
amplitude at zero energy
fixed at its experimental value 
by adjusting $2\mu R^2 V_{\rm S}(R)$, which
displays an periodic dependence on a power of the 
cutoff~\cite{Beane:2000wh,Bawin:2003dm,Braaten:2004pg,PhysRevA.71.022108,
Hammer:2005sa,PavonValderrama:2007nu,Bouaziz:2014wxa,Odell:2019wjq}.
For $n=2$, 
the dependence is periodic in $\ln R$, characteristic of a limit cycle
and a remaining {\it discrete} scale invariance.
(For discussions of limit cycles, see
Refs. \cite{Hammer:2011kg,Bulycheva:2014twa}.)
The $n\ge 3$ oscillation indicates a generalized limit cycle.
The case $n=4$ is displayed in 
Fig. \ref{contactforr-2} \cite{Beane:2000wh}.

Having renormalized zero-energy scattering, 
an important question is whether the problem is well defined also at 
finite energy $E\equiv k^2/(2\mu)$. 
That this is the case can be shown \cite{Beane:2000wh}
with the WKB approximation, which applies
to the region where the wavelength is small
compared to the characteristic distance over which the potential varies
appreciably. For distances where $|V_{\rm L}(r)|\gg E$,
one recovers Eq. \eqref{wfsingpot} for the wavefunction,
up to energy-dependent
corrections that are determined by Eq. \eqref{wfsingpot} itself. 
In the absence of a short-range interaction, decrease
in $R$ would lead to the repeated appearance of low-energy bound states
due to the unstoppable growth in attraction,
a phenomenon reflected in the never-ending oscillations of 
the wavefunction \cite{Perelomov:1970fz}.
With $V_{\rm S}(R)$ preventing this collapse
and ensuring the description of {\it one} low-energy datum,
bound states can accrete only from negative energies,
converging to finite values as $R$ decreases.
How many of the bound states are within the region of 
validity of the EFT depends, of course, on the scales in the problem:
the very low-energy spectrum will be affected by the long-distance tail
of the potential while states with binding energies 
$\simge (2\mu R_{\rm und}^2)^{-1}$ are irrelevant for the distances of interest.
For $n=2$ and $\alpha >\alpha_{0}$, 
which is equivalent \cite{Efimov:1971zz} to the 
three-boson system with short-range interactions at unitarity, 
the bound states form a geometric tower (``Efimov states'' \cite{Efimov:1970zz})
that signals the remaining discrete scale invariance
stemming from the limit cycle in the contact interaction
\cite{Bedaque:1998kg,Bedaque:1998km}.
While the existence of the tower is a consequence of the symmetry,
its position is fixed by the LEC.
It is remarkable that it is the proper renormalization of 
the EFT that underlies the ``Efimov physics'' intensely 
explored with cold atoms
\cite{Braaten:2004rn}.

A particularly simple example of singular potential is the delta function
itself. In this case the external potential vanishes
and the external zero-energy wavefunction 
is replaced by
\begin{equation}
\psi_0(r)\propto r^{-1} \left( 1 -\frac{r}{a_0} +\ldots \right)\, ,
\label{nopot}
\end{equation}
where $a_0$ determines the ratio between irregular and regular solutions
and is nothing but the scattering length. 
The solution for Eq. \eqref{match} can be written explicitly,
\begin{equation}
V_{\rm S}(R)= -\frac{1}{2\mu  R^2}
\left[\left(1+2m\right)^2\frac{\pi^2}{4}+\frac{2R}{a_0}+\ldots \right] \, ,
\label{deltaren}
\end{equation}
where $m$ is an integer.
It is apparent how a cutoff-dependent $C_0(R)\propto R$
softens the delta function.
The scattering length enters in the smaller $R^{2}$ term.
Of course, a similar result is obtained for a momentum cutoff 
$\Lambda \sim R^{-1}$ \cite{vanKolck:1998bw}.

A subtlety arises when a regular potential with $n=1$ in 
Eq. \eqref{singpot} is present 
together with the delta function, as is the case for OPE.
By itself, the long-range potential needs no regularization;
with the delta function, a new cutoff dependence emerges
in the irregular solution \cite{Beane:2001bc,PavonValderrama:2007nu}:
\begin{equation}
\psi_0(r)\propto r^{-1} \left\{ 1 
- r \left[\frac{1}{\overline{a}_0}
+ \alpha \left(\ln\frac{r}{R_\star}-1\right) \right]+\ldots \right\}\, ,
\label{Yukout}
\end{equation}
where $\overline{a}_0$ and $R_\star$ are length scales that enter
the zero-energy scattering amplitude.
Instead of Eq. \eqref{deltaren},
\begin{equation}
V_{\rm S}(R)= -\frac{1}{2\mu  R^2}
\left[\left(1+2m\right)^2\frac{\pi^2}{4}
+2R \left(\frac{1}{\overline{a}_0} + \alpha \, \ln\frac{R}{R_\star}\right) 
+\ldots \right] 
\, .
\label{deltaYukren}
\end{equation}
The main difference is the appearance of the $\ln R$ with a coefficient
$\propto \alpha$. 

In both these cases, where the outside potential is not singular, 
it is easy to see that the amplitude at finite energy is well defined. 
The energy
enters both internal and external wavefunctions as $(kr)^2$
and can only affect
the depth of the spherical well by a term of ${\cal O}(R^0)$, an effect
that disappears as $R$ decreases. 
The multiple branches in Eqs. \eqref{deltaren} 
and \eqref{deltaYukren} are a consequence of the fact that
a spherical well can have multiple bound states.
The zero-energy amplitude is essentially determined
by the shallowest state, and we can choose different well
depths to place any one state at the desired position.
Deeper states have energies $\propto (2\mu R^2)^{-1}$ and, again,
are beyond the regime of the EFT for $R\simle R_{\rm und}$.
Differently from long-range singular potentials,
the three-dimensional delta function supports a single
bound or virtual state.

\subsubsection{Higher partial waves}
\label{singnonperthigher}

We can now look at higher partial waves. Amplitudes in these waves
have additional powers of $\vec{p}\,'\cdot \vec{p}$, 
where $\vec{p}$ ($\vec{p}\,'$)
is the relative incoming (outgoing) nucleon momentum.
Just as for $k^2$ in the $S$ wave,
in the absence of a long-range potential,
dimensional analysis implies that $\vec{p}\,'\cdot \vec{p}$
must come together with $R^2$:
the no-derivative contact interaction contributes in the 
small-$R$ limit only to the $S$ wave. 
For the $n=1$ external potential, the $l\ge 1$ phase shifts
then converge as $R\to 0$.
A long-range singular potential
of the type \eqref{singpot} contains an intrinsic scale $r_0$
and $\vec{p}\,'\cdot \vec{p}$ comes in general with a factor $r_0^2$
and does not disappear as $R\to 0$. There is a phase $\phi_l$ in
Eq. \eqref{wfsingpot} for every $l$, which can only be fixed
by higher-derivative interactions.

To see this, let us first stick to the potential \eqref{singpotcomplete}.
The $k=0$ matching equation that generalizes Eq. \eqref{match} is
\begin{equation}
R_l(R)\equiv \sqrt{-2\mu R^2 V_{\rm S}(R)} \;
\frac{j_{l+1}(\sqrt{-2\mu R^2 V_{\rm S}(R)})}
     {j_{l}(\sqrt{-2\mu R^2 V_{\rm S}(R)})}
=l+1 
-\left.r\frac{\partial}{\partial r} \ln \left(r \psi_l(r)\right)\right|_{r=R} 
\,,
\label{matchl}
\end{equation}
where $j_l$ is the spherical Bessel function of the first kind. 
Using the recurrence relation for Bessel functions,
\begin{equation}
R_l(R) = 2l+1 +\frac{2\mu R^2 V_{\rm S}(R)}{R_{l-1}(R)}\,.
\label{recurrence}
\end{equation}

In the absence of an external potential, the external wavefunction
is a combination of the regular $j_l$ and the irregular $y_l$,
the spherical Bessel function of the second kind. 
By direct calculation we find that at small $R$ 
\begin{equation}
R_l(R) = 2l+1 +{\cal O}(R^{2l+1}/a_l)\,,
\label{Rltozero}
\end{equation}
where $a_l$ is the $l$-wave scattering ``length'' ({\it e.g.} volume
for $l=1$), the zero-energy limit
of the ratio of the $y_l$ and $j_l$ coefficients.
Using $R_0(0)=1$ in Eq. \eqref{recurrence} gives
\begin{equation}
R_1(0)=3 - \left[(2n+1)\frac{\pi}{2}\right]^2 \,.
\end{equation}
which implies, together with Eq. \eqref{Rltozero}, that 
$a_1={\cal O}(R^{3})$. The argument repeats for $l\ge 2$ with
different finite pieces, leading to $a_l={\cal O}(R^{2l+1})$. 
As anticipated by dimensional analysis,
the effect of the non-derivative contact interaction disappears from
$l\ge 1$ waves as $R\to 0$.
A similar argument for a regular outside potential leads to the
same conclusion. For the argument with a delta-shell regularization,
see Ref. \cite{PavonValderrama:2010fb}.

In contrast, 
when the external potential is attractive and singular with $n\ge 3$, 
\begin{equation}
R_l(R) = l+1 -\frac{n}{4}+ \sqrt{\alpha} \, R^{1-n/2}
\tan\left(\frac{\sqrt{\alpha}}{n/2-1}R^{1-n/2}+\phi_l\right) \,.
\label{Rlsingpot}
\end{equation}
Matching in the $S$ wave makes $\phi_0$ $R$-independent.
Since $2\mu R^2V_{\rm S}(R)$ is approximately cutoff independent
as can be seen from either of the two approximate solutions 
\eqref{matchlarge} and \eqref{matchsmall}, 
Eq. \eqref{recurrence} gives 
\begin{equation}
R_1(R)= 3 - \Delta_1(R) \,,
\end{equation}
where $\Delta_1(R\ll r_0)$ is finite.
Comparison with Eq. \eqref{Rlsingpot} then shows that
$\phi_1\propto R^{1-n/2}$. Continuing to larger $l$ we 
find 
\begin{equation}
\phi_l (R\ll r_0)= -\frac{\sqrt{\alpha}}{n/2-1}R^{1-n/2} \,.
\label{phasel}
\end{equation}
The phases are thus angular-momentum and energy independent 
\cite{PavonValderrama:2010fb} in this limit, 
{\it but cutoff dependent} \cite{Perelomov:1970fz}.

What is needed for renormalization is a single contact interaction with a 
minimum number of derivatives in each wave, with LECs $C_{2l}'$. 
The interaction is non-local, for example for $l=1$,
\begin{equation}
\frac{C_{2}'}{4\pi r^2} \left(\frac{\partial\delta(r)}{\partial r}\right)
\left.\frac{\partial}{\partial r'}\right|_{r'=0}
\to \frac{C_{2}'(R)}{4\pi R^3}\left[\frac{2}{r}\,\theta(R-r)-\delta(r-R)\right]
\left.\frac{\partial}{\partial r'}\right|_{r'=R}\, ,
\label{contact1}
\end{equation}
where $C_{2}'(R)$ is determined so as to keep the
phase $\phi_1$, and thus one $P$-wave low-energy datum, fixed. 
The contact interactions 
are all determined by the underlying interactions,
but without additional dynamical assumptions we do not know
how they relate to each other. 
Model independence requires we keep them free.

\subsubsection{Implications}
\label{singnonpertmoral}

Much of the above had been understood without EFT.
The use of boundary conditions, for example, goes back at least
to the work of Breit \cite{Breit:1947zza}.
In EFT, a boundary condition corresponds to a specific 
regulator. At the two-body level, in the $S$ wave
we have simply traded the dependence in $R$ by that of $V_{\rm S}(R)$.
Renormalization means that, as far as observables are concerned, 
the regulator choice is irrelevant (within the error of the truncation);
only the unobservable cutoff dependence of the LECs depends on the regulator.
What matters is that a LEC encodes one parameter.
The LO EFT in coordinate space is in the spirit of 
atomic Quantum-Defect Theory, where the interaction of far-away electrons 
with an ionic core or molecule is solved for exactly and
a few parameters (``defects'') account for short-range interactions
\cite{Greene:1982}.

The model independence of the EFT is manifest 
in the fact that the same two-body contact interactions that renormalize the
two-body problem contribute to other processes. 
For example, 
the three-boson system was considered in Ref. \cite{Odell:2019wjq},
where binding energies and the particle-dimer scattering length 
were calculated.
Convergence was observed in a range of cutoffs, with
asymptotic values representing model-independent predictions.
The role of $D$ and higher waves in these results was,
however, not discussed.

The contact interactions can also be seen as providing a self-adjoint 
extension of the Hamiltonian.
As stressed in Ref. \cite{Behncke:1968}, 
the so-called deficiency index for a singular potential is 
$(\infty,\infty)$, that is, an infinite
number of parameters --- the phases $\phi_l$ in Eq. \eqref{wfsingpot} 
for all values of $l$ --- are needed to determine the self-adjoint 
extension uniquely.
In the EFT this translates into the existence
of an infinite number of contact interactions, one with
the minimal number of derivatives for each wave.
(Of course, the EFT contains also contact interactions with
an arbitrary number of derivatives.)

While mathematically the problem looks hopeless, 
on physical grounds this is clearly a red herring.
As remarked in Ref. \cite{Nogga:2005hy},
increasing $l$ strengthens the centrifugal barrier
and shrinks the distances $r\simle [l(l+1)]^{-1/(n-2)}r_0$
where the attractive $n\ge 3$ potential takes over.
The distance of closest approach at momentum $k$ can be estimated from
the point where the energy is comparable to the centrifugal barrier,
or $r\simge [l(l+1)]^{1/2} k^{-1}$.
For $k\simle M_{\rm und}$, the breakdown scale,
we are only interested in distances 
$r\simge R_{\rm und}\sim [l(l+1)]^{1/2} M_{\rm und}^{-1}$.
We might then expect that only in waves with $l\simle l_{\text{cr}}$ 
does a singular potential need to be treated exactly
and Eq. \eqref{wfsingpot} apply, where \cite{Birse:2005um}
\begin{equation}
l_{\text{cr}}\left(l_{\text{cr}}+1\right)\sim \frac{r_0}{R_{\rm und}} \,.
\label{lcr}
\end{equation}
A more precise semi-analytical estimate comes from
the investigation of the critical strength $\alpha$
where a 
Bessel series solution of the Schr\"odinger equation exhibits
a square-root branch point characteristic of 
nonperturbative behavior.
For $n=3$ \cite{Birse:2005um},
it is described pretty well for large $l$ by the estimates above.
For $n=2$, consideration of the first two orders in the perturbative expansion 
suggests $l_{\rm cr}=(\pi |\alpha| -2)/4$ \cite{Long:2007vp}.
An attractive singular potential defined with a step function at 
$l_{\text{cr}}$ has a finite deficiency index $(l_{\text{cr}},l_{\text{cr}})$.

The situation is different in the case of $n=1$.
The potential is larger than both centrifugal barrier and kinetic repulsion for
$r\simge n^2(l)\, r_0$, where $n(l)$ is ${\cal O}(1)$ for $l=0$ and 
grows as $l$ for large $l$.
Balance among these terms leads to bound states of sizes
$r_n\sim 2 n^2 r_0$ and 
binding energies $B_n \sim \alpha^2/(8\mu n^2)$. 
(Taking as an example the Coulomb interaction, 
where $\alpha=2\mu \alpha_{\rm e}$ in terms of the 
fine-structure constant $\alpha_{\rm e}$, 
we get the proper result $B\sim \mu \alpha_{\rm e}^2/(2 n^2)$
if we interpret $n$ as the principal quantum number.)
These estimates are in any case affected by the long-range tail
of the potential, which we are not considering in this section.
But at distances $R_{\rm und}\simle r\simle r_0$, we expect
$l_{\rm cr}\approx 1$: while the $S$ wave might
be nonperturbative and perhaps require a short-range potential 
\eqref{deltaYukren} to generate a bound state at the observed location,
higher waves should be perturbative.

\subsection{Perturbative corrections}
\label{singpert}

EFT provides a framework where we can systematically incorporate
corrections to the leading interactions, which can be checked
with the method developed in Ref. \cite{Griesshammer:2015osb}.
We pair subleading long-range interactions with the subleading short-range
interactions needed for renormalization order by order.
As stressed in Ref. \cite{Long:2012ve},
renormalization at a given order contains clues about the relative
importance of higher corrections. 
Just as a negative power of $R$ indicates at least one missing LECs, 
so positive powers of $R$ point to
the order before at least one new LEC should appear.
If the error in an observable 
not used in the fit of LECs at N$^{i}$LO (with some integer $i$)
scales as a positive power of the coordinate cutoff, say $R^x$, 
then we may expect that corrections
appear at N$^{i+j}$LO, where $j\le x$ is an integer
(not necessarily the largest integer).
This constraint comes from the demand that the regulator error
should not exceed the truncation error when $R\simle R_{\rm und}$.
(It does not exclude the presence of a LEC at a lower order than 
that estimated by the cutoff dependence,
corresponding to boundary conditions of the RG equation 
\cite{Valderrama:2014vra}).
We will see examples below.

The next renormalization challenge arises from
the more-singular corrections to the long-range potential.
An almost automatic reflex is to simply add the correction to the LO potential,
as Weinberg prescribed,
and solve the Schr\"odinger equation. 
For a regular potential, adding a regular correction
that is small everywhere can be done in perturbation theory,
but it can also be done by solving the 
Schr\"odinger equation exactly. For a more-singular correction, however,
the perturbing potential will be larger than the LO potential at 
sufficiently small $r$. 
One risks destroying the systematic character of the EFT
unless one keeps $R$ relatively large.
Whether this risk materializes needs to be checked explicitly.
As we will see, renormalization
requires distorted-wave perturbation theory around the LO solution
\cite{Nogga:2005hy,Long:2007vp}.
Implications for nuclear interactions are discussed in 
Sec. \ref{partpert}.

\subsubsection{Distorted-wave perturbation}
\label{singpertdistorted}

A pedagogical toy model that nicely illustrates the need for perturbation
theory on singular corrections was presented in Ref. \cite{Epelbaum:2009sd}.
The model consists of two separable, regular potentials,
one of range $m_{\rm L}^{-1}$, the other of range $m_{\rm S}^{-1}\ll m_{\rm L}^{-1}$.
Because the potentials are separable, exact answers can be found
for the effective-range parameters.
The potential parameters are fine-tuned so that each potential separately
produces a natural scattering length, that is, $a_0\sim m_{\rm L}^{-1}$
($a_0\sim m_{\rm S}^{-1}$) in the absence of the short-range (long-range)
potential. 
Next, the short-range potential is expanded in powers of $k/m_{\rm S}$, 
creating a series of singular interactions.
While for $k\sim m_{\rm L}$ the long-range 
potential is nonperturbative, the singular corrections
should be treated in distorted-wave perturbation theory.
Lo and behold, the results up to N$^2$LO 
obtained with a standard subtraction scheme
are found to reproduce the exact results.
In contrast, when a truncation of the expanded short-range potential
is solved exactly, similar to the ``peratization'' of Fermi 
theory~\cite{Feinberg:1963zz,Feinberg:1964zza}, 
one can no longer take a large momentum cutoff.
Reference \cite{Epelbaum:2009sd} concludes that removing the cutoff dependence 
is impossible, which is indeed true when one insists on iterating subleading 
corrections.

The situation is not significantly different for the case of interest in
nuclear physics where not only corrections, but also
the LO potential is singular.
Again, the simplest example is provided by the delta function without external
potential, $V_{\rm L}(R)=0$ in Eq. \eqref{singpotcomplete}.
As discussed above, the energy dependence first affects the matching
between internal and external wavefunctions at relative ${\cal O}(k^2R^2)$.
The ratio of irregular and regular solutions, which determines 
$k\cot\delta_0(k)$ where $\delta_0(k)$ is the $S$-wave phase shift, 
starts at ${\cal O}(R)$.
Thus, at LO
\begin{equation}
k\cot\delta_0(k)= -\frac{1}{a_0} \left(1+ {\cal O}(Ra_0k^2)\right) \,,
\label{kcotdeltafordelta}
\end{equation}
which means that the fractional error in $\delta_0$ is
\begin{equation}
\frac{\Delta\delta_0(k)}{\delta_0(k)}= {\cal O}(Ra_0k^2) \,.
\label{errorfordelta}
\end{equation}
For example, the effective range $r_0\sim R$. This again can be easily 
obtained with a momentum regulator \cite{vanKolck:1998bw}.
In ChEFT, where away from the chiral limit the delta function is 
accompanied in the singlet $S$ wave by the Yukawa potential,
the situation is not substantially different \cite{PavonValderrama:2007nu}.
Aside the ${\cal O}(\alpha R \ln R)$ dependence in Eq. \eqref{deltaYukren},
the argument does not change and Eq. \eqref{errorfordelta} still holds
with $a_0\to {\overline a}_0$.
Despite the presence of pions, the error is still $\propto R$.
It can be removed in first-order
perturbation theory by a two-derivative contact interaction
\begin{equation}
\delta V_{\rm S}=C_2\left\{\left[\nabla^2\delta(\vec{r})\right] 
+2\left[\vec{\nabla}\delta(\vec{r})\right] \cdot \vec{\nabla}
+2\delta(\vec{r})\nabla^2\right\}\, ,
\label{contact2}
\end{equation}
whose LEC $C_2(R)\propto R^2$ fixes $r_0\sim R_{\rm und}$.
For $R\simle R_{\rm und}$, this contact interaction
is an NLO correction to the LO interaction with LEC $C_0$. 
This is in fact one of the elements
in the power counting in Pionless EFT \cite{Hammer:2019poc}.
Note that, if we were to impose that $C_2/C_0$ scaled with $R_{\rm und}^2$ 
as implied by NDA,
we would obtain an effective range
that scaled the same way, in contrast to what one obtains for 
typical short-range potentials \cite{vanKolck:1998bw}.
Once again, renormalization automatically enforces a general property
of short-range interactions.

But what if we solved the Schr\"odinger equation exactly following
Weinberg's prescription? In the simpler case without
a long-range potential, it has been shown explicitly 
\cite{Phillips:1996ae,Phillips:1997xu,Beane:1997pk} 
that this can be done in a renormalized way only if $r_0\simle R$,
which is arbitrarily small.
In other words, the two-derivative contact
interaction is nonperturbatively renormalizable only if the theory satisfies 
a ``Wigner bound'' \cite{Wigner:1955zz} $r_0\ge 0$.
In contrast, when the two-derivative contact
interaction is treated in perturbation theory, 
at second order and higher, which contain loops involving two or
more powers of $C_2$, 
four- and higher-derivative contact interactions
appear to guarantee renormalization. 
When we resum the two-derivative contact
interaction we generate diagrams with an arbitrary number of loops,
but lack the counterterms to remove the cutoff dependence.
A calculator committed to exact solutions
might be tempted to eschew renormalization (and thus model independence)
and live with a relatively large $R$. 
Still, such stubbornness in resumming what needs no resummation
might be rewarded by results that are {\it worse} than those
of the perturbative expansion.
An example is provided by a calculation \cite{Stetcu:2010xq}
of the $S$-wave scattering phase shifts for a 
harmonically trapped unitary system,
where the regulator was implemented in the form of a maximum number of shells.
One can see explicitly how in first-order
perturbation theory the derivatives in Eq. \eqref{contact2}
give a contribution to the NLO energy which 
is proportional to the LO energy, apart from a shift
in the LO LEC. The result of resumming the NLO interaction is not
only cutoff dependent but also gives rise to  
a larger violation of unitarity than even NLO.

Note that one can introduce an auxiliary ``dimeron'' field
in the EFT Lagrangian \cite{Kaplan:1996nv} whose kinetic term provides 
an energy-dependent correction to the potential.
Exploiting the redundancy of interactions in the enlarged Lagrangian,
one can remove the momentum-dependent corrections \eqref{contact2}.
Renormalization changes with an energy-dependent potential 
and, in particular, a resummation does not restrict $r_0$.
However, unless there is evidence for $r_0\gg R_{\rm und}$, this is still 
an NLO correction and the resummation does not affect observables up 
to higher-order terms \cite{vanKolck:1998bw}.

Resummation of subleading interactions can lead to an even more
paradoxical situation. 
The problem is that subleading singular 
potentials are not in general attractive 
in all the same waves as OPE.
If the corrections are iterated together with OPE, the cutoff behavior
of the amplitude will change completely:
channels that required a counterterm at LO may not require, or even tolerate, 
one at subleading order
\cite{PavonValderrama:2007nu}.
Take a wave where the LO potential is singular with a power $n$
and attractive, thus
requiring a counterterm, but the subleading potential is repulsive
(strength $\alpha'$)
with a power $n'>n$.
The exact solution of the Schr\"odinger equation for the sum
of the external potentials
is now dominated at short distances by the irregular 
solution of the subleading potential,
which grows exponentially as $r$ decreases.
Matching to the short-range potential $V_{\rm S}$ will force a non-vanishing
irregular solution, which in turn leads to an
exponentially {\it increasing} dependence
of the fractional phase shift error in $R$, 
$\propto R^{1+n'/2} \exp[2 \sqrt{-\alpha'} \, R^{1-n'/2}/(n'/2-1)]$
\cite{PavonValderrama:2007nu}.
The only solution is to remove the LO LEC at subleading order!
There is hardly a way to keep the systematic expansion of the EFT.

Another toy model \cite{Epelbaum:2018zli} 
illustrates this paradox. 
This time the underlying potential consists
of a {\it repulsive} $r^{-3}$ component 
associated with a mass $m_{\rm L}$ 
together with an attractive $r^{-3}$ from a heavier $m_{\rm S}\gg m_{\rm L}$,
as well as less singular terms.
Its exact $S$-wave results are compared to those of a potential
consisting of the repulsive $r^{-3}$ potential plus a delta-function interaction.
Parameters are chosen so that the repulsive potential 
is nonperturbative. Despite the fact that the phase shifts 
of the repulsive component are well defined by themselves, 
Ref. \cite{Epelbaum:2018zli} includes the delta function nonperturbatively,
fixing it to reproduce the scattering length of the underlying potential.
For $R^{-1}\simle m_{\rm S}$ the phase shifts are in reasonably good agreement 
with those of the underlying potential. 
However, agreement deteriorates as $R$ decreases.
Disregarding conceptual differences in renormalization
of attractive and repulsive singular potentials 
\cite{Beane:2000wh,PavonValderrama:2007nu},
Ref. \cite{Epelbaum:2018zli} concludes that cutoff dependence cannot
be removed in general, rather than
in the particular case of resumming the subleading
delta function. 
In response, Ref. \cite{Valderrama:2019yiv} included 
the $2n$-derivative delta functions, which account for the short-range 
potential, at N$^{2(n+1)}$LO in perturbation theory.
Calculations up to N$^{8}$LO show convergence to the exact
phase shifts up to at least $k\sim 2m_{\rm L}$
without significant restriction on $R$.
(Reference \cite{Epelbaum:2019msl} nevertheless points to
some ambiguity in the values of the NLO phase shifts, apparently
implying that it is sufficient reason to abandon renormalization.)

Thus the singular nature of the potentials that we want to treat
in an EFT expansion of the amplitude 
requires the use of perturbation theory on corrections,
as implied by the power counting of Sec. \ref{lessons}.
This in fact ensures small changes in amplitudes after
renormalization \cite{Long:2007vp}.
But then one might wonder to which extent the singular nature
of the LO potential affects the order of the corrections. 
As we have seen, when the only singular part of the LO potential 
is a delta function, the first correction comes at NLO.
When the outside potential is singular and attractive, 
the situation is different.
For an LO singular attraction, one finds \cite{PavonValderrama:2007nu}
that after fixing the phase $\phi_0$ the $S$-wave phase shifts scale as
\begin{equation}
\frac{\Delta\delta_0(k)}{\delta_0(k)}\propto R^{1+n/2} \,.
\label{errorforsingattractive}
\end{equation}
This means that corrections are expected at (or before) N$^2$LO for
$n=2,3$, N$^3$LO for $n=4,5$, {\it etc.}. 
It is unclear why the results reported in Ref. \cite{Odell:2019wjq}
indicate higher sensitivity to $R$ than given by 
Eq. \eqref{errorforsingattractive}.

Now, the power counting for nuclear interactions in Sec. \ref{lessons}
says that at N$^2$LO there are corrections to the long-range potential
with an $r^{-(n+2)}$ singularity. 
The additional singularity can be removed in first-order
perturbation theory by additional contact interactions with two derivatives.
This can be shown relatively simply in a toy
model where a $\pm r^{-4}$ potential is added to
an $n=2$ attractive LO potential \cite{Long:2007vp}.
The analysis was carried out in momentum space with a sharp cutoff $\Lambda$.
At N$^2$LO, where the $\pm r^{-4}$ potential is considered as a first-order
perturbation, two forms of additional, oscillating cutoff dependence
appear: one proportional to $\Lambda^2$, reflecting the
stronger singularity of the perturbing potential, the other proportional 
to $k^2$. In the $S$-wave, a two-derivative potential 
\eqref{contact2} is sufficient,
together with an N$^2$LO shift in the $C_0$ of Eq. \eqref{contact}, 
to remove the two additional divergences.
This argument can presumably be continued at higher orders
and repeated for $l\ge 1$ waves by considering
interactions of type \eqref{contact1} with two more derivatives.
One tentatively concludes that NDA holds in distorted-wave perturbation
once it has been corrected at LO.

\subsubsection{Simple perturbation}
\label{singpertsimple}

In partial waves $l\simge l_{\rm cr}$ where the LO potential is perturbative
and particles are free in zeroth approximation,
corrections are included in simple perturbation theory.
The first task in this case is to quantify the angular-momentum suppression
for the long-range potentials so as to establish the orders they come in.
The second need is to find the orders the associated contact interactions
appear at.

For the $\mu=0$ long-range potential, rules \eqref{red1} and \eqref{red2}
indicate that a contact interaction is needed for renormalization
at $n$th order in perturbation theory, where $n\ge 2l+1$.
This is consistent with the inference from
the residual cutoff dependence of the non-derivative contact interaction.
As we saw in Sec. \ref{singnonperthigher},
$l$-wave scattering ``lengths'' $a_l$ are induced
through matching at finite $R$.
Just as for the $S$-wave effective range, 
they can be made arbitrarily small by taking $R\to 0$.
However,
the higher power of $R$, $R^{2l+1}$, suggests that contact interactions
in higher waves enter in perturbation theory
at N$^{2l+1}$LO or lower, another element of
Pionless EFT power counting \cite{Hammer:2019poc}.

The increased singularity of subleading potentials
asks for counterterms at lower orders in perturbation theory. 
The first-order perturbative correction due to subleading potentials 
involving pion loops is renormalized with LECs assigned by NDA.
Making further general statements about the order contact interactions
are needed is cumbersome without an explicit angular-momentum
suppression factor.

If one were to solve the Schr\"odinger equation exactly in 
one of these waves, renormalization would require a LEC, which then determines
the asymptotic properties of the wavefunction. 
The tail of the nonperturbative wavefunction can be
reproduced with increasing accuracy as the
order of perturbation theory increases \cite{Beane:2000wh}. 
Being a series in $\alpha$, the perturbative expansion cannot
reproduce the oscillations found in Eq. \eqref{wfsingpot},
which are tied to the non-analytic dependence $\sqrt{\alpha}$.
This is no problem because,
by definition of $l_{\rm cr}$, these oscillations take place at distances
smaller than those probed by the EFT.
Their effects can be ``averaged out'' and appear through
contact interactions at subleading orders. 
If one wants to save all the perturbative work 
by sticking to a nonperturbative solution,
one loses some predictive power at LO but, because it is
a single LEC (in one wave), this is perhaps acceptable.
Alternatively, one could simply not include the LEC if $l$ is
sufficiently high for oscillations to happen below $R$,
which might be limited in numerical calculations anyway.
In this case $R$ is in the region where perturbation theory works
and the result will be relatively insensitive to $R$.
Unnecessary iteration in high waves
is thus relatively harmless, other than obscuring the systematic EFT expansion.

\section{Renormalization of Chiral EFT}
\label{RenormChi}

By this point in the manuscript
it should be clear how to proceed with ChEFT in the nuclear sector.
The power counting of ChPT is based on NDA, which comes from
demanding that the EFT expansion be renormalized order by order
so as to ensure model independence.
In the more general ChEFT we continue to insist on model independence,
but now LO is nonperturbative.
The results of the previous section apply to pion-exchange 
potentials, where the spin-isospin factors and the exponential
fall-off at large $m_\pi r$ do not substantially affect renormalization.
Perhaps not surprisingly in hindsight, NDA is violated.

Since the OPE tensor force is singular and attractive in an infinite
number of channels, 
the first task (Sec. \ref{partpert}) is to estimate up to which 
relative angular momentum $l$ OPE needs to be considered at LO.
In Secs. \ref{chiEFT2N} and \ref{chiEFTmoreN}
renormalized results for, respectively, two and more nucleons are described.

\subsection{Partly perturbative pions}
\label{partpert}

The simple power counting of Eqs. \eqref{red1} and \eqref{red2}
does not capture factors of $l^{-1}$, just as it misses other dimensionless
factors. 
More realistically, OPE in the radial
Schr\"odinger equation is an expansion in $Q/M_{N\!N}^{(l,s)}$, where 
$M_{N\!N}^{(0,s)} \sim M_{N\!N}$ but $M_{N\!N}^{(l,s)}$ increases with $l$
depending in general also on the spin $s$.
Once $M_{N\!N}^{(l_{\rm cr}^{(s)},s)}\sim \MQCD$, OPE is perturbative.
What do we know about $M_{N\!N}^{(l,s)}$ and $l_{\text{cr}}^{(s)}$?

The bold suggestion was made in Refs. \cite{Kaplan:1998tg,Kaplan:1998we}
that $l_{\text{cr}}^{(s)}\approx 0$, so that
pion exchange would be amenable to perturbation theory in {\it all} waves.  
The estimate in Eq. \eqref{OPEscale} assumed NDA for the one-nucleon quantities
$m_N={\cal O}(\MQCD)$, $f_\pi={\cal O}(\MQCD/(4\pi))$, and
$g_A= {\cal O}(1)$, plus neglected
any dimensionless factors. 
Numerically, $M_{N\!N}\simeq 290$ MeV.
What if the various spin/isospin factors and other numbers floating around, 
each of ${\cal O}(1)$, conspire to make OPE more perturbative,
so that $M_{N\!N}$ is effectively comparable to $\MQCD$?

In that case, at LO ChEFT would be formally the same as 
Pionless EFT \cite{Bedaque:2002mn,Hammer:2019poc}, 
where the binding of light nuclei rests on the shoulders of
non-derivative $2N$ and $3N$ contact interactions 
\cite{Bedaque:1999ve,Konig:2016utl}.
But because pions are explicit, the range of validity of the EFT is enlarged
--- at least near the chiral limit where integrating out pions becomes a very 
restrictive condition.  
An attractive feature of this proposal 
is that it could potentially explain why Pionless EFT works 
better than expected, for example for binding energies \cite{Hammer:2019poc}.

This proposal also neatly solves the renormalization issues of the last section.
OPE is now an NLO effect of relative ${\cal O}(Q/M_{N\!N})$,
so no problems associated with its singularity emerge.
Being perturbative, it brings NLO cutoff dependence only to $S$ waves. 
Because at LO the external potential vanishes,
Eq. \eqref{kcotdeltafordelta} 
requires at NLO one chirally symmetric
two-derivative contact interaction in each $S$ wave.
Then $Q\sim m_\pi$ implies the concomitant presence
of a chiral-symmetry-breaking non-derivative interaction with
LEC proportional to the quark masses, $m_\pi^2 D_2$.
In the background of an LO wavefunction
of the type \eqref{nopot}, OPE generates an $m_\pi^2\ln \Lambda$
cutoff dependence which can be absorbed in $D_2$.
The $2N$ amplitude is renormalized and in good agreement
\cite{Kaplan:1998tg,Kaplan:1998we,Soto:2007pg}
with the Nijmegen partial-wave analysis (PWA) \cite{Stoks:1993tb}
up to $Q\sim m_\pi$.

\begin{figure}[t]
\begin{center}
\includegraphics[scale=.65]{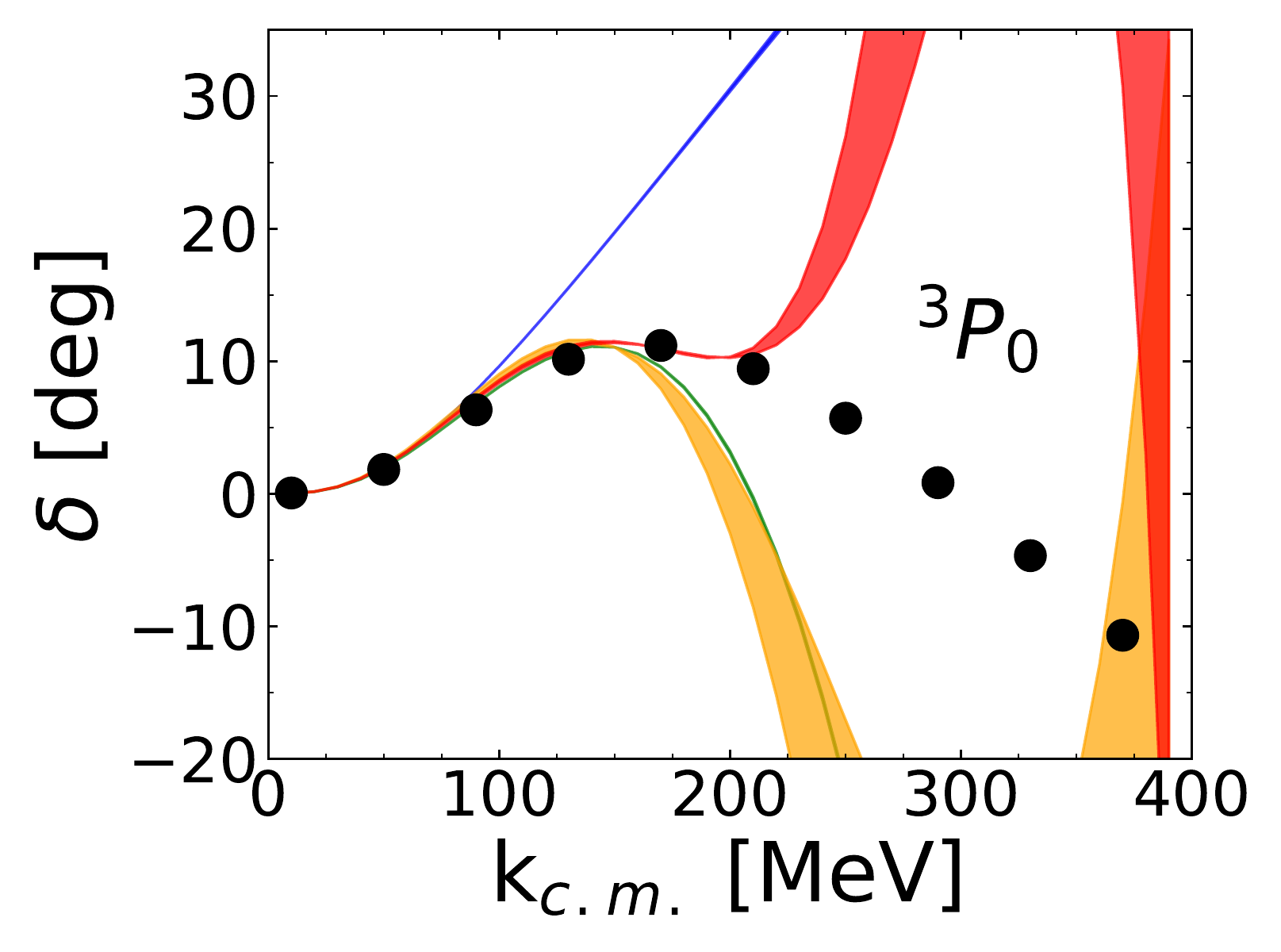}
\end{center}
\caption{Two-nucleon $^3P_0$ phase shift $\delta$ 
as function of the center-of-mass momentum $k_{\rm c.m.}$. 
The NLO (blue), N$^2$LO (green), N$^3$LO (orange), and N$^4$LO (red) bands 
from a perturbative treatment of pion exchange 
correspond to cutoff variation from 0.8 to 2.4 GeV.
(LO in a perturbative expansion vanishes for this channel.)
The empirical phase shifts from the SAID program~\cite{SAID} (solid circles)
are shown for comparison.
Reprinted figure with permission from Ref. \cite{Wu:2018lai}.
Copyright (2019) by the American Physical Society.}
\label{pertpions}
\end{figure}

Alas, calculations at ${\cal O}(Q^2/M_{N\!N}^2)$ have shown 
\cite{Cohen:1999ds,Fleming:1999ee} that in 
the low, spin-triplet partial waves,
where the OPE tensor force is attractive, the 
expansion fails for $Q\sim 100$ MeV.
In partial waves with $l=j \gg 1$, 
where counterterms are needed only at a very large
number of loops $L\ge 2l$, the breakdown of perturbation
theory was estimated in the chiral limit to be at a 
critical momentum \cite{Kaplan:2019znu}
\begin{equation}
p_{\rm cr}\approx \frac{l^3}{\sqrt{27}|2(-1)^l+1|} M_{N\!N} \, .
\label{s=1l=jpcr}
\end{equation}
If we impose $p_{\rm cr}\sim \MQCD$,
we get $l_{\text{cr}}^{(1)}\approx 2.5$.
The radius of convergence of the perturbative series is
not as large in waves with $l=j\pm 1$. 
In both cases the first few orders were found \cite{Kaplan:2019znu}
not to be representative of the large-order convergence.
For low partial waves counterterms enter already at low orders. 
When they were assigned arbitrary but natural values,
all waves except $^3S_1$-$^3D_1$, $^3P_0$, and perhaps $^3P_1$ were found to 
converge up to $p_{\rm cr}\approx M_{N\!N}$.
An example
of failure, $^3P_0$, is given in Fig. \ref{pertpions} \cite{Wu:2018lai},
where OPE is NLO, $n$-iterated OPE N$^n$LO, 
leading two-pion exchange (TPE) N$^3$LO,
and subleading TPE N$^4$LO.
The LECs are assumed to be given by NDA 
instead of being introduced only at the order where they are 
first needed for renormalization.
These signs of the breakdown of 
perturbative pions are consistent with an expansion in $Q/M_{N\!N}^{(l,1)}$ with 
$M_{N\!N}^{(l\approx 1,1)}\sim f_\pi$ as indicated by NDA.

It seems inevitable that pions must be treated nonperturbatively
in the low partial waves if we want to go beyond Pionless EFT
at physical quark masses.
Still, based on the general arguments of Sec. \ref{singnonpertmoral}
we expect pions to be perturbative for sufficiently high partial waves.
The $n=3$ tensor force, for which $r_0 \sim M_{N\!N}^{-1}$,
does not vanish for spin $s=1$.
Equation \eqref{lcr} with $R_{\rm und}\sim [l(l+1)]^{1/2} M_{\rm QCD}^{-1}$ 
provides an estimate
$l_{\text{cr}}^{(1)}\approx 2$ for the critical angular momentum
in attractive triplet waves.
This conclusion is made firmer by a generalization to the tensor potential
of the analysis of the onset of square-root branch points in the 
Bessel series solution of the Schr\"odinger equation \cite{Birse:2005um}.
Given that the strength of OPE is fixed by $M_{N\!N}$, 
it translates into an upper bound on the critical momentum
$p_{\text{cr}}$, including repulsive waves. 
The results, listed in Table~\ref{OPEtensorcrit},
are obtained in the chiral limit; a realistic pion mass
could affect the smaller values by factors of ${\cal O}(1)$ but
is not expected to be important for the larger values.
They indicate that OPE in
$^3S_1$-$^3D_1$ and $^3P_0$ likely fails to converge already below $M_{N\!N}$.
In contrast, OPE in high waves, such as $F$ and higher, converges
beyond $\MQCD$.
The gray zone is the $D$ and $P$ waves other than $^3P_0$.
Given the low values of $p_{\text{cr}}$ on the scale set by $\MQCD$,
one might conclude that $l_{\text{cr}}^{(1)}\approx 3$.
An analysis of spin-triplet
phase shifts where OPE and TPE are removed 
in distorted-wave perturbation \cite{Birse:2007sx}
supports this conclusion.

\begin{table}[tb]
\begin{center}
\begin{tabular}{|c|c|}
\hline
Partial wave & $p_{\text{cr}}$/MeV\\
\hline
\hline
$^3S_1$-$^3D_1$ & 66 \\
$^3P_0$ & 182 \\ 
$^3P_1$ & 365 \\
$^3P_2$-$^3F_2$ & 470 \\ 
$^3D_2$ & 403 \\ 
$^3D_3$-$^3G_3$ & 382 \\
$^3F_3$ & 2860 \\ 
$^3F_4$-$^3H_4$ & 2330 \\ 
$^3G_4$ & 1870 \\
\hline
\end{tabular}
\end{center}
\caption{Estimate of the critical values $p_{\text{cr}}$ of the 
relative momentum in the lowest 
$2N$ triplet channels above which the OPE tensor force cannot be treated 
perturbatively~\cite{Birse:2005um}.
\label{OPEtensorcrit}}
\end{table}

A different but closely related estimate for $l_{\text{cr}}^{(1)}$
comes from the cutoff values where the first bound state
crosses threshold in the absence of contact interactions.
The very early work on ChEFT and much of its phenomenological
improvements, which continue to this day, have used Weinberg's prescription.
Unfortunately this prescription assigns to triplet waves a single 
non-derivative contact interaction at LO,
which is incapable to determine more than one phase in a model-independent way.
In particular, for a separable regulator the contact interaction 
contributes only to the $S$ wave. Spurious low-energy bound states can be 
kept at bay at LO in the $^3S_1$-$^3D_1$ coupled channel
\cite{Frederico:1999ps,Beane:2001bc,PavonValderrama:2005gu,Yang:2007hb,Song:2016ale}, 
but only in this channel \cite{Nogga:2005hy,PavonValderrama:2005uj}.
In triplet waves where OPE is repulsive there is no
need for counterterms at LO~\cite{Eiras:2001hu,Nogga:2005hy},
but without them bound states 
repeatedly cross threshold in attractive waves and 
lead to wild variations in the phase shifts at energies
within the realm of ChEFT 
\cite{Nogga:2005hy,PavonValderrama:2005uj,Epelbaum:2006pt,Song:2016ale}.
With a super-Gaussian separable regulator, bound states first emerge at,
roughly, 
$\Lambda \sim$ 0.5, 1, 2, 4, and 6 GeV in respectively
$^3S_1$-$^3D_1$, $^3P_0$, $^3D_2$, $^3P_2$-$^3F_2$, and $^3D_3$-$^3G_3$ 
channels \cite{Nogga:2005hy,Song:2016ale}.
Except for $^3D_3$-$^3G_3$, this sequence is similar to that of the attractive
channels in Table~\ref{OPEtensorcrit}.
The lowest two channels would display shallow states when
$\Lambda \sim \MQCD$, indicating that OPE is
nonperturbative, while the higher waves are less clear
--- numerical experimentation suggested \cite{Nogga:2005hy}
their effects were not 
negligible, which can be understood from the results of
Ref. \cite{Birse:2005um}.

Perhaps even more seriously, in Weinberg's scheme
more-pion exchange and
other contact interactions, which should be treated perturbatively,
are not. This leads to the pathologies discussed in Sec. \ref{singpert}.
Indeed, renormalization problems have been 
reported~\cite{PavonValderrama:2005wv,PavonValderrama:2005uj,Entem:2007jg,
Yang:2009kx,Yang:2009pn,Zeoli:2012bi}
within Weinberg's prescription also for higher-order potentials.
These renormalization failures prevent taking a momentum-space cutoff 
at the breakdown scale $\MQCD$ or higher.
A ``physical cutoff'' $\Lambda_{\text{phys}}\simle 1$ GeV, 
before $^3P_0$ would develop a bound state~\cite{Nogga:2005hy}, is needed,
and results are sensitive to the choice of regulator.
No wonder then that much effort in phenomenology
with chiral potentials has been dedicated to finding the ``best'' regulator.
The limitation to small cutoffs leads to startling dependence
on what should be equivalent forms of interactions in the Lagrangian,
see for example Ref. \cite{Lynn:2015jua}.

One concludes that, while it seems well established that to handle
triplet waves beyond $M_{N\!N}$ pions are nonperturbative in
at least $^3S_1$-$^3D_1$ and $^3P_0$,
there is some uncertainty as to the partial wave up to which
this is so.
Part of the uncertainty comes from the presence of LECs
in lowest orders of the amplitude, which require a closer comparison
with data (Sec. \ref{chiEFT2N}).
What is clear is that there is an angular-momentum suppression.
The perturbative expressions in Ref. \cite{Kaplan:2019znu} suggest
\begin{equation}
M_{N\!N}^{(l,1)}\sim l^2 M_{N\!N} \, ,
\label{MNNs=1}
\end{equation}
apart from an overall suppression of $l^2$.
In contrast, the analyses of Ref. \cite{Birse:2005um} leads
to $l^2\to [l(l+1)]^{3/2}$.

Singlet channels are somewhat simpler, but not devoid of subtleties.
Since the tensor force vanishes for $s=0$, OPE has $n=1$ and 
$r_0 \sim M_{N\!N}/m_\pi^{2}$.
The general argument from Sec. \ref{singnonpertmoral}
indicates that only in the $S$ wave should we expect nonperturbative
effects, $l_{\text{cr}}^{(0)}\approx 1$. 
In higher waves, the OPE potential dominates over 
kinetic and centrifugal repulsion only at large distances,
and there the exponential fall-off of OPE leads to further
suppression.

The perturbative convergence of the $l\ge 1$ channels was studied
in Ref. \cite{PavonValderrama:2016lqn}. This is particularly easy
because the Yukawa potential is well defined for an
arbitrary number of loops. The phase shifts are seen to converge
quickly already for $^1P_1$, and faster as $l$ increases.
The suppression factor $M_{N\!N}^{(l,0)}$ can be estimated from the 
critical strength
$M_{N\!N\text{cr}}^{-1}$
needed to generate a zero-energy bound state in the corresponding $l$ wave,
shown in Table \ref{OPEsingcrit}. 
There are two sequences of channels
that alternate because of the factor of $-3$ in
the ratio between isospin singlet and triplet:
if we multiply the isosinglet entries in Table \ref{OPEsingcrit}
the results form a single monotonous sequence.
Assuming $Q\sim m_\pi$, we find that in each sequence increasing $l$ by 2
roughly suppresses OPE by one order in the expansion, starting with $^1P_1$
at NLO and $^1D_2$ at N$^2$LO.
Moreover,
\begin{equation}
M_{N\!N}^{(l,0)}\sim \left[ l(l+1) +1 \right] M_{N\!N} \, ,
\label{MNNs=0}
\end{equation}
in the isosinglet waves, with a factor 3 larger in isotriplets.

\begin{table}[tb]
\begin{center}
\begin{tabular}{|c|c|}
\hline
Partial wave & $M_{N\!N}/M_{N\!N\text{cr}}$\\
\hline
\hline
$^1P_1$ & $-6.4$ \\
$^1D_2$ & $45.8$  \\ 
$^1F_3$ & $-27.9$ \\ 
$^1G_4$ & $133.1$  \\ 
$^1H_5$ & $-64.6$ \\ 
$^1I_6$ & $265.9$ \\ 
$^1J_7$ & $-116.4$ \\ 
$^1K_8$ & $440.0$ \\ 
$^1L_9$ & $-183.3$ \\ 
$^1M_{10}$ & $667.4$ \\ 
$^1N_{11}$ & $-265.4$ \\ 
\hline
\end{tabular}
\end{center}
\caption{Estimate of the critical strength $M_{N\!N\text{cr}}^{-1}$ of the 
Yukawa potential in the lowest 
$2N$ singlet channels above which OPE cannot be treated 
perturbatively~\cite{PavonValderrama:2016lqn}.
\label{OPEsingcrit}}
\end{table}

If one insists on the full solution for the Yukawa potential in higher partial 
waves, there are no renormalization problems \cite{Eiras:2001hu,Nogga:2005hy}, 
as the potential is regular.
In the $S$ wave, however, interference with the delta function
leads to an unexpected violation of NDA. As first noticed in 
Ref. \cite{Kaplan:1996xu}
and confirmed many times since --- for example, 
Refs. \cite{Beane:2001bc,PavonValderrama:2003np,PavonValderrama:2005wv}---
cutoff dependence proportional to $m_\pi^2$ 
emerges through the $\ln R$ term in Eq. \eqref{deltaYukren}.
Renormalization therefore requires 
the non-derivative chiral-symmetry-breaking interaction with LEC
$m_\pi^2 D_2$.
With Weinberg's prescription, where this LEC is missed at LO,
the cutoff dependence can be 
seen in the $2N$ system only if quark masses are varied,
as one does to match lattice QCD results.
From the perspective of phenomenology, the main effect of the absence of
the $m_\pi^2 D_2$ contact interaction is in processes
sensitive to its associated pion interactions,
which are generated by the way chiral symmetry is broken explicitly in QCD.
Regardless of its
phenomenological (ir)relevance, this is the simplest example where the 
renormalization of observables in ChEFT is not guaranteed by NDA.  

Clearly, dimensionless factors stemming from spin and isospin
make the transition from nonperturbative to perturbative OPE
somewhat fuzzy. Moreover, virtually nothing has been done to estimate the
angular-momentum suppression for multiple-pion exchange.
Multiple-pion exchange is amenable to perturbation theory in all waves, 
but presumably further suppressed in higher waves. 
That is sufficient to start comparing with data.

\subsection{Two Nucleons}
\label{chiEFT2N}

Let us now take a closer look at how a renormalized approach works at
the $2N$ level. I continue to consider $Q \sim m_\pi \sim M_{N\!N}$.
Since the OPE tensor force survives in the chiral limit, if we take
$m_\pi \simle M_{N\!N}$ we can perform an additional expansion around the chiral 
limit~\cite{Beane:2001bc}, but such an expansion in $m_\pi/M_{N\!N}$ 
has not been fully explored.

Leading order at the $2N$ level 
consists of the exact solution
of the Schr\"odinger equation up to $l_{\rm cr}^{(s)}$
with OPE and the required counterterms,
not all of which were accounted for by NDA:
\begin{itemize}
\item Two non-derivative, chirally symmetric contact interactions with LECs 
$C_{0(s)}$,
one for each $S$ wave ($s=0,1$). They are needed to renormalize
OPE even in the chiral limit, and were anticipated 
\cite{Weinberg:1990rz,Weinberg:1991um} to appear at LO already
on the basis of NDA, which estimates $C_{0(s)}\sim 4\pi/(m_N M_{N\!N})$.

\item A non-derivative, chiral-symmetry-breaking contact interaction
with LEC $m_\pi^2 D_{2(0)}$ 
if OPE is treated nonperturbatively in the $^1S_0$ channel. 
This LEC is $D_{2(0)}\sim C_{0(0)}/M_{\rm QCD}^2$ on the basis
of NDA, and thus N$^2$LO.
Renormalization of nonperturbative OPE 
instead requires $D_{2(0)}\sim C_{0(0)}/M_{N\!N}^2$ \cite{Kaplan:1996xu}. 

\item One chirally symmetric contact interaction with the minimum number 
of derivatives for each wave where attractive tensor OPE is iterated.
The most dramatic effect is in $^3P_0$, 
where a contact interaction $C'_{2(1)} \vec{p}\,'\cdot \vec{p}$
with $C'_{2(1)}\sim C_{0(1)}/M_{N\!N}^2$ is needed \cite{Nogga:2005hy}. 
NDA would give instead $C'_{2(1)}\sim C_{0(1)}/M_{\rm QCD}^2$.
The two-order enhancement comes from the running of pion exchange,
and similarly enhancements apply for the LECs in other attractive,
singular waves where OPE is nonperturbative.  
\end{itemize}
These counterterms are schematically displayed in Table \ref{contactorders},
assuming $l_{\rm cr}^{(1)}=3$.

\begin{table}[tb]
\begin{center}
\begin{tabular}{|c||c|c|c|c|c|c|c|}
\hline
& $^1S_0$ & $^3S_1$ & $\epsilon_1$ & $^3P_0$, $^3P_2$ & $^1P_1$, $^3P_1$ 
& $\epsilon_2$ & $^3D_2$, $^3D_3$\\
\hline
\hline
LO      & 1 
        & 1 
        & 
        & $p'p$ 
        & 
        & 
        & $p'^{\,2}p^{2}$ 
\\
NLO     & $p'^{\,2}+p^{2}$ 
        & 
        &
        &
        & 
        & 
        & 
\\ 
N$^2$LO & $p'^{\,4}+p^{4}$ 
        & $p'^{\,2}+p^{2}$ 
        & $p^2$
        & $p'p \left(p'^{\,2}+p^{2}\right)$ 
        & $p'p$ 
        & $p'p \, p^2$ 
        & $p'^{\,2}p^{2}\left(p'^{\,2}+p^{2}\right)$ 
\\
N$^3$LO & $p'^{\,6}+p^{6}$ 
        & 
        &
        &
        &
        & 
        & 
\\ 
\hline
\end{tabular}
\end{center}
\caption{Schematic momentum dependence of the lowest-order contact
interactions in the $2N$ system up to $D$ waves, according to Refs. 
\cite{Beane:2001bc,Nogga:2005hy,Long:2007vp,Long:2011qx,Long:2011xw,Long:2012ve}.
\label{contactorders}}
\end{table}

\begin{figure}[t]
\begin{center}
\includegraphics[scale=.5]{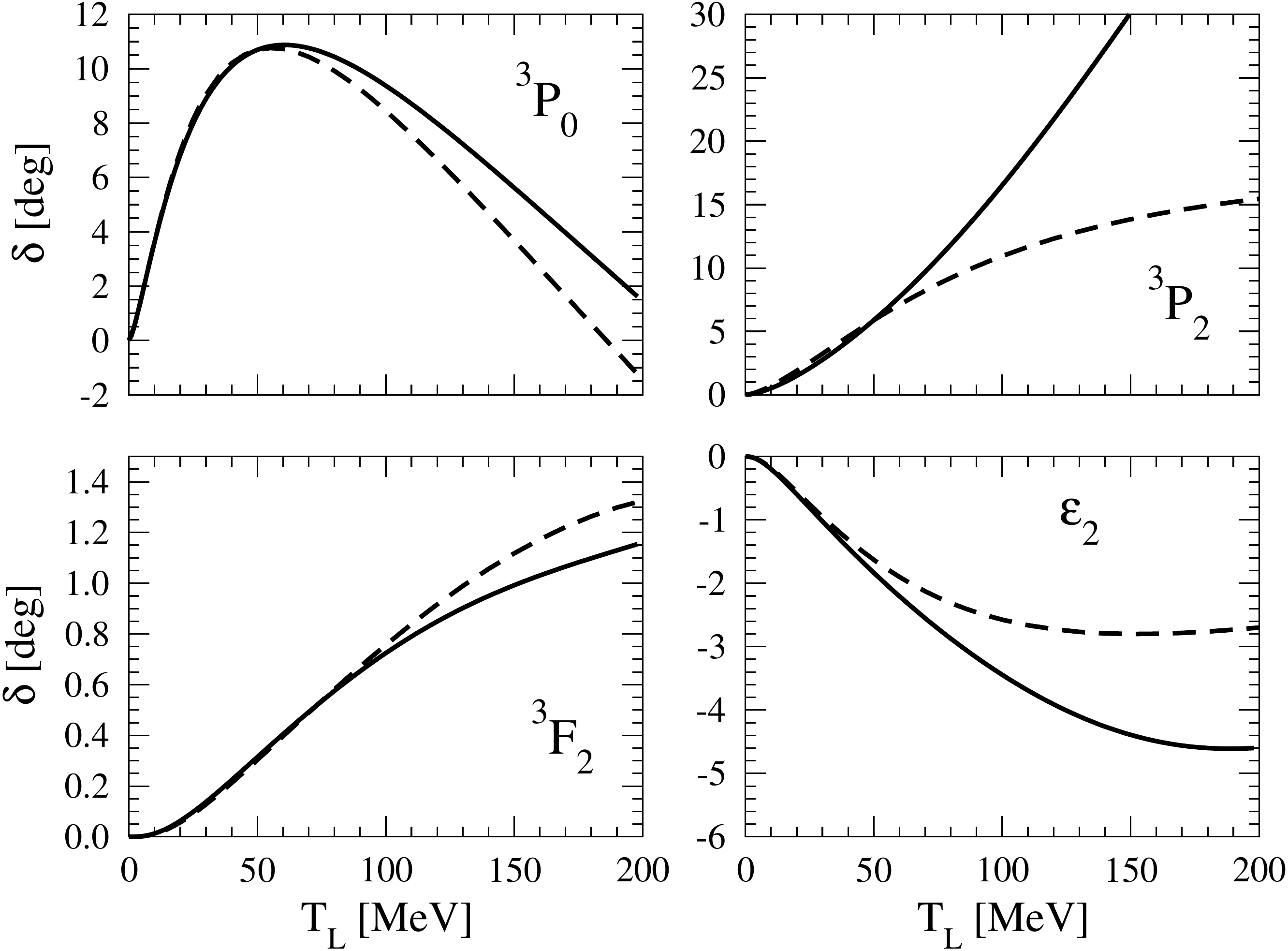}
\end{center}
\caption{
Two-nucleon $^3P_0$ and $^3P_2$-$^3F_2$ phase shifts ($\delta$) and 
mixing angle ($\varepsilon_2$) as functions of the laboratory energy $T_{\rm L}$.
The LO results (solid lines) at a cutoff $\Lambda=3.94$ GeV
are compared with the 
Nijmegen PWA \cite{Stoks:1993tb} (dashed lines).
Reprinted figure with permission from Ref. \cite{Nogga:2005hy}.
Copyright (2005) by the American Physical Society.}
\label{renormChEFTLO}
\end{figure}

Results can be found in Refs. 
\cite{Beane:2001bc,Nogga:2005hy,Epelbaum:2006pt,Song:2016ale} 
for cutoff
values as high as 10 GeV in super-Gaussian separable regulators.
In comparison with the Nijmegen PWA, one finds:
\begin{itemize}
\item
In the $^3S_1$-$^3D_1$ coupled channels, where Weinberg's prescription
is consistent with renormalization,
phase shifts come out well with one fitted LEC.
Results improve for $\Lambda \simge \MQCD$;
even the mixing angle, which is somewhat 
overpredicted with a small $\Lambda\sim 500$ MeV, agrees with the Nijmegen PWA
to within $1^\circ$ up to a laboratory
energy $E_{\rm lab}\simeq 200$ MeV for $\Lambda\simge 4$ GeV.
When the scattering length is used to fix the LEC,
the deuteron binding energy is $B_2^{\rm LO}\simeq 2.0$ MeV,
which is essentially the same as for lower cutoffs \cite{Lynn:2017fxg}.

\item
For low uncoupled, attractive triplet channels ($^3P_0$, $^3D_2$)
iterating pions with one fitted LEC works equally well.
As an example, Fig. \ref{renormChEFTLO} \cite{Nogga:2005hy}
shows $^3P_0$, which comes out much better than in Weinberg's prescription
with $\Lambda\sim 500$ MeV.
(Compare this also with Fig. \ref{pertpions} where pions are treated 
perturbatively.)
The vanishing of the amplitude beyond 
$E_{\rm lab}\simeq 200$ MeV can be described, because attraction from
OPE is compensated by the contact interaction.
Again, agreement improves with increasing cutoff.

\item
For low coupled triplet channels ($^3P_2$-$^3F_2$, $^3D_3$-$^3G_3$)
--- see Fig. \ref{renormChEFTLO} \cite{Nogga:2005hy} again for an example ---
iterated pions
with the associated LEC do not improve significantly
over Weinberg's prescription with $\Lambda\sim 500$ MeV.
While $^3D_3$ is much better, changing from repulsion
to attraction, $^3P_2$ goes from underprediction to considerable
overprediction.

\item
In triplet channels without free parameters 
($^3P_1$, $^3F_3$, $^3F_4$-$^3H_4$, $^3G_4$)
iterated pions tend to work well, whether they are
expected to be perturbative or not.
In these channels results are the same as in Weinberg's prescription;
there is not much change as $\Lambda \simge \MQCD$.

\item
In $l\ge 1$ singlet channels ($^1P_1$, $^1D_2$, $^1F_3$, $^1G_4$), 
iterated pions undershoot data except in $^1F_3$.
Again results essentially agree with Weinberg's prescription
at small $\Lambda\sim 500$ MeV. 

\item
In $^1S_0$, the phase shifts resemble those of Pionless EFT, where after the
fast rise due to the existence of a virtual state, they remain essentially
flat as $E_{\rm lab}$ increases. 
Weinberg's prescription applies, and renormalization allows us to increase
the cutoff beyond $\MQCD$, but agreement with the Nijmegen PWA
deteriorates as we do so.
\end{itemize}
Thus, a renormalized approach where the regulator is unimportant
gives a qualitative guide to $2N$ data at LO,
which is slightly better than Weinberg's prescription with specific 
regulators and small momentum-cutoff parameters. 
It has been shown recently \cite{Tews:2018sbi} that, 
with a non-separable regulator,
a specific combination of the four possible spin-isospin non-derivative
contact interactions that yields only one $^3S_1$-$^3D_1$ bound state
simultaneously prevents bound states in other channels.
While this is not true for an arbitrary regulator, it does allow to
extend LO results with Weinberg's prescription
to higher cutoff values, in general improving agreement with the
Nijmegen PWA. However, results are not clearly better than the renormalized
approach, particularly in the $^3P_0$ channel which 
lacks the repulsion to produce the amplitude zero.

In addition to simple perturbative corrections
in higher partial waves, one needs to account in subleading orders
for potential corrections {\it via} distorted-wave perturbation theory
in the lowest partial waves. 
The residual $\Lambda^{-1}$ dependence of the LO amplitude
means that at NLO --- relative ${\cal O}(Q/\MQCD)$ ---
there is also:
\begin{itemize}
\item A two-derivative, chirally symmetric contact interaction with LEC 
$C_{2(0)}$ in the $^1S_0$ channel. 
In order to render cutoff effects on the effective range 
no larger than N$^2$LO, $C_{2(0)}\sim C_{0(0)}/(M_{N\!N}M_{\rm QCD})$
\cite{Long:2012ve}.
NDA gives instead $C_{2(0)}\sim C_{0(0)}/M_{\rm QCD}^2$, or N$^2$LO
(confusingly denoted NLO in the nuclear community),
which produces a short-range contribution to the effective range
smaller than pion's by two powers of the expansion parameter.
Yet, only about half of the $^1S_0$ effective range comes from OPE.
\end{itemize}
The cutoff dependence in other channels is milder,
in agreement with the discussion of Sec. \ref{singpert}.
The NLO interaction is shown in the second line
of Table \ref{contactorders}.
At NLO in the amplitude,
the NLO interaction should be included in first order in the distorted-wave
expansion.

At higher orders, corrections to the long-range potential
enter according to the power counting of Sec. \ref{lessons}.
Barring unforeseen renormalization issues, at ${\cal O}(Q^\mu/\MQCD^\mu)$ 
we need to include LECs with 
up to $\mu$ derivatives more than the LECs appearing at LO~\cite{Long:2007vp}, 
except in the $^1S_0$ channel where
the Yukawa/delta-function interference takes place.
The momentum structures of the LECs up to N$^3$LO are shown in 
Table \ref{contactorders}, again under the assumption
$l_{\rm cr}^{(1)}=3$. They are:
\begin{itemize}
\item 
In each triplet channel where attractive OPE is iterated at LO
($^3S_1$-$^3D_1$, $^3P_0$, {\it etc.}),
a contact interaction with two derivatives
more than the contact interaction at LO 
\cite{Long:2011qx,Long:2011xw}. 
While for $^3S_1$-$^3D_1$ this coincides with NDA,
for other channels NDA would say these contact interactions only appear
at N$^4$LO or higher.

\item 
Contact interactions with two derivatives \cite{Long:2012ve} for 
singlet ($^1P_1$) and triplet $P$ waves 
where OPE is repulsive ($^3P_1$).
This is the NDA scaling.

\item 
Four- and six-derivative contact interactions in the $^1S_0$ channel
at N$^2$LO and N$^3$LO, respectively \cite{Long:2012ve}.
Again, NDA would have these contact interactions at N$^4$LO or higher.
\end{itemize}
Up to N$^3$LO in the amplitude,
their contributions are included in first order in the distorted-wave
expansion.
Meanwhile, the NLO interaction must be included in second and 
third orders, either by itself or with one N$^2$LO interaction.

\begin{figure}[t]
\begin{center}
\includegraphics[scale=.4]{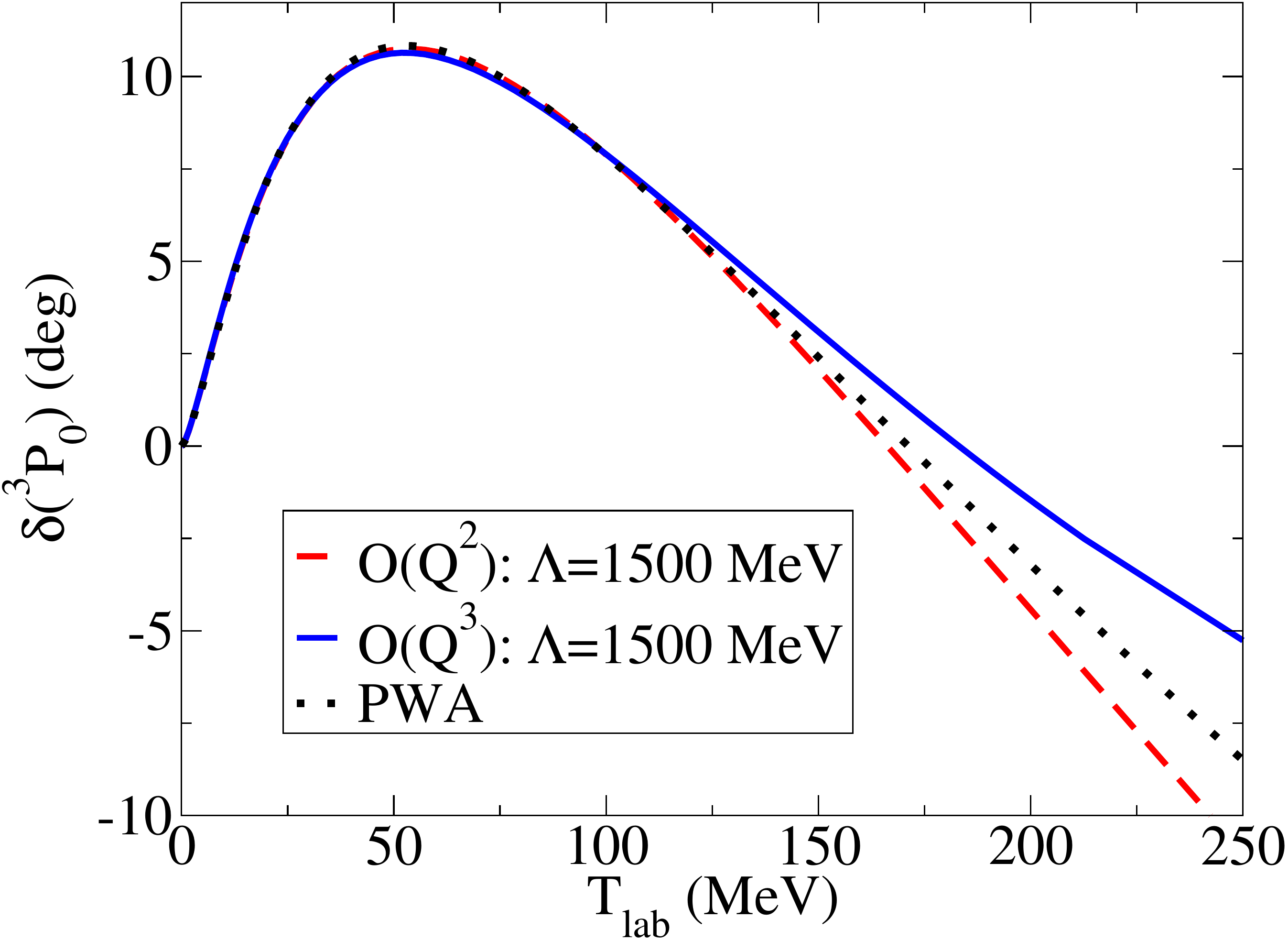}
\end{center}
\caption{Two-nucleon $^3P_0$ phase shift $\delta(^3P_0)$ as function
of the laboratory energy $T_{\rm lab}$.
The N$^2$LO (red dashed line) and N$^3$LO (blue solid line) results at a cutoff 
$\Lambda =1.5$ GeV are compared with the Nijmegen PWA 
\cite{Stoks:1993tb} (black points).
Reprinted figure with permission from Ref. \cite{Long:2011qx}.
Copyright (2011) by the American Physical Society.}
\label{LongYang}
\end{figure}

The phase shifts have been calculated up to N$^3$LO
along these lines in Refs. 
\cite{Long:2011qx,Long:2011xw,Long:2012ve}, along with Delta{\it less} TPE:
\begin{itemize}
\item
In the $^3S_1$-$^3D_1$ coupled channels,
where LO already yielded very good results at LO,
results improve only marginally at N$^{2,3}$LO.

\item
In $^3P_0$, which was also relatively well described at LO,
results improve quite a bit around the maximum phase shift at N$^2$LO.
Not much improvement, if any, is seen at N$^3$LO.
Results from Ref. \cite{Long:2011qx} are shown in Fig. \ref{LongYang},
to be compared with LO in Ref. \ref{renormChEFTLO}.
Other uncoupled, attractive triplet channels ($^3D_2$ {\it etc.})
were not calculated.

\item
The coupled $^3P_2$-$^3F_2$ wave with OPE iterated at LO 
shows no real improvement
at N$^2$LO, and only mildly better agreement with the Nijmegen PWA at N$^3$LO.
No results are available for higher coupled triplet channels
($^3D_3$-$^3G_3$ {\it etc.}).

\item
In $^3P_1$, which works well at LO with no free parameter,
results deteriorate at N$^{2,3}$LO.
Higher repulsive triplet channels
($^3F_3$ {\it etc.}) were not considered.

\item In $^1P_1$, agreement with the Nijmegen PWA improves at N$^{2,3}$LO, 
although results are very sensitive to the pion-nucleon
parameters that enter the $\mu=3$ TPE.
Higher singlet partial waves were not studied.

\item The $^1S_0$ phase shift improves considerably at NLO
but is still not very close to the Nijmegen PWA. N$^2$LO improves further,
but the zero of the amplitude is still poorly described.
\end{itemize}
Overall, there is some improvement at N$^{2}$LO but not much
at N$^{3}$LO. This is perhaps an indication that a better description of the 
pion-nucleon subamplitude with an explicit Delta isobar is needed.

Note that subleading corrections have also been calculated in Refs. 
\cite{Valderrama:2009ei,Valderrama:2011mv} with a slightly different
accounting of higher orders. 
For example, TPE is taken to start
three orders higher than OPE, which is contrary to the power counting of 
Sec. \ref{lessons} and difficult to conciliate with the power counting
used in ChPT.
Still, results are generically not much different from those described above.
A third power-counting
variant has been proposed \cite{Birse:2005um} with similar features.
It has not been tested in detail,
perhaps because no clear prescription is given
for handling the LO cutoff dependence in a channel like $^3P_0$
where a counterterm is assigned relative ${\cal O}(Q^{1/2}/\MQCD^{1/2})$.
Reference \cite{Griesshammer:2015osb} discusses these alternatives.

\begin{figure}
  \centering
  \includegraphics[scale=0.337]{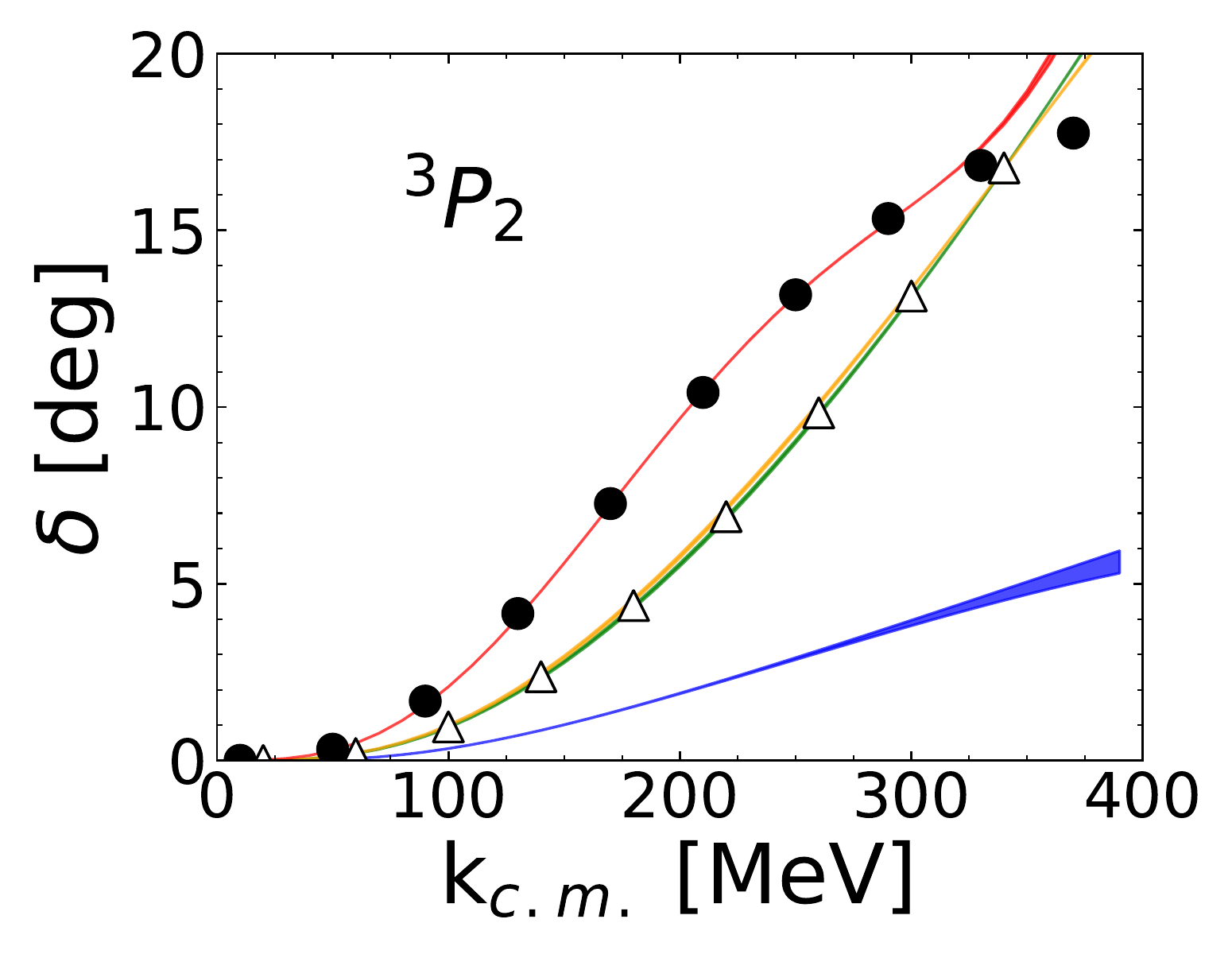}
  \includegraphics[scale=0.337]{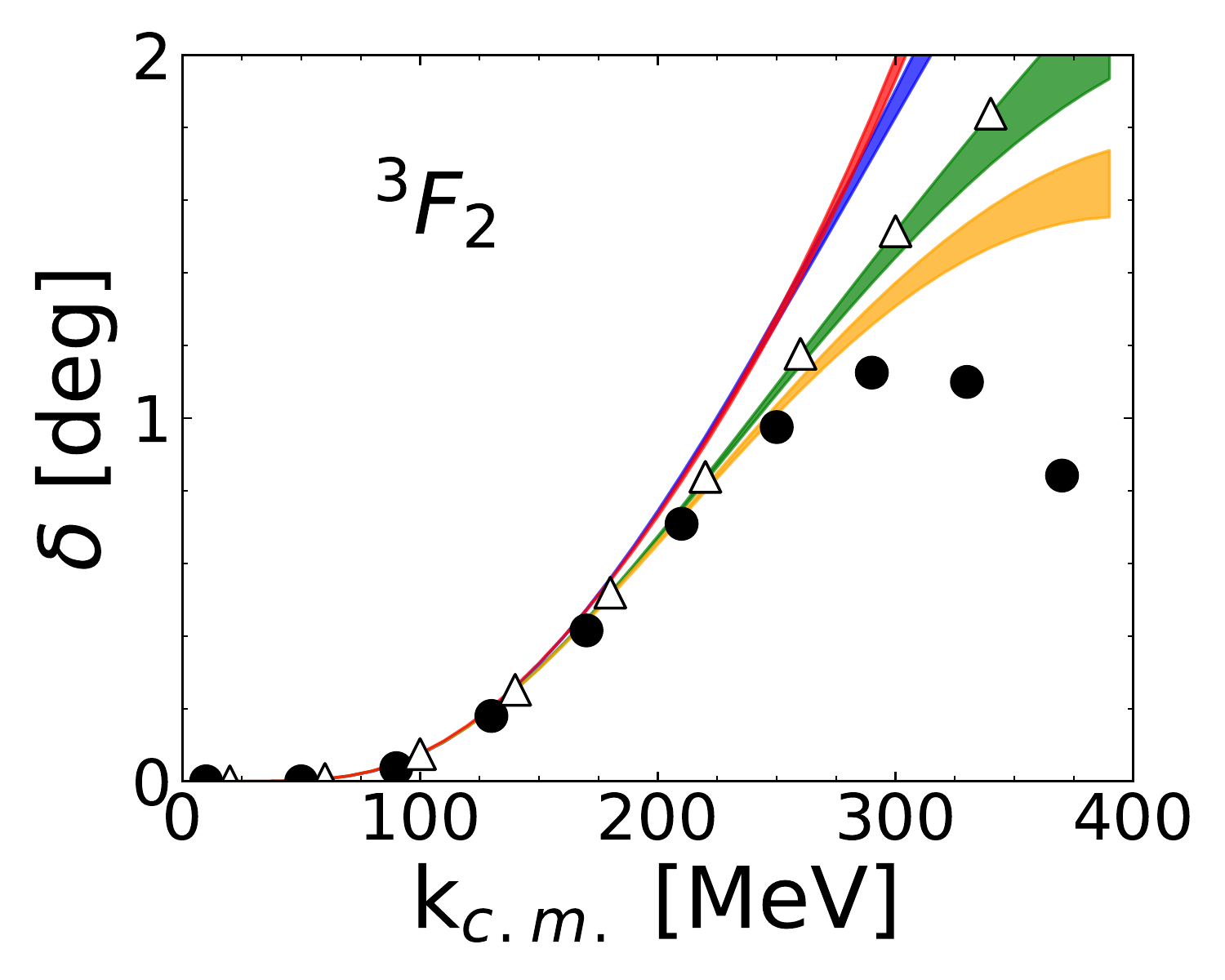}
  \includegraphics[scale=0.337]{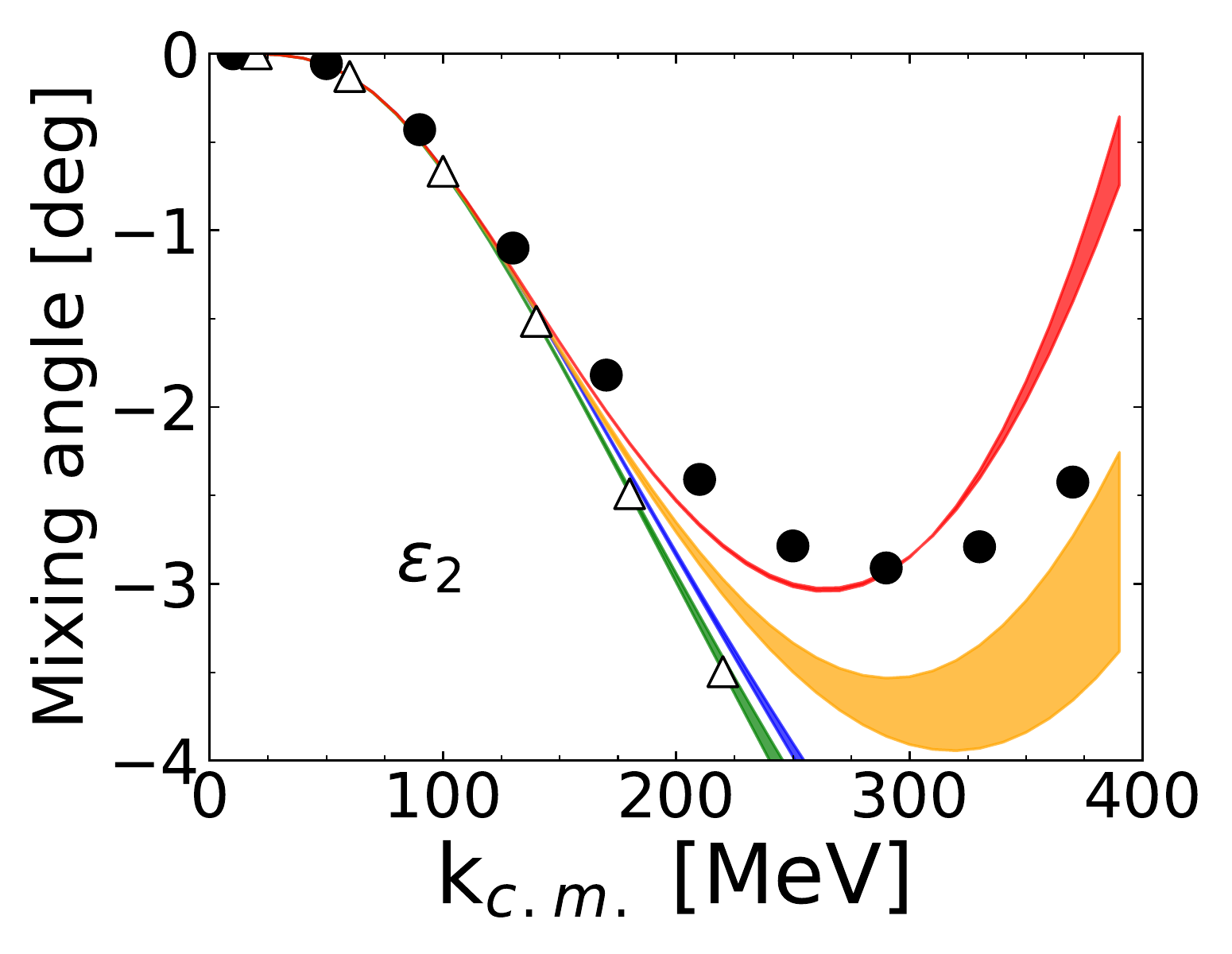}
  \caption{Two-nucleon $^3P_2$-$^3F_2$ phase shifts ($\delta$) and 
mixing angle ($\varepsilon_2$)
as functions of the center-of-mass momentum $k_{\rm c.m.}$. 
The NLO (blue), N$^2$LO (green), N$^3$LO (orange), and N$^4$LO (red) bands 
from a perturbative treatment of pion exchange 
correspond to cutoff variation from 0.8 to 4.8 GeV.
N$^2$LO results for $\Lambda \to \infty$ are also shown (triangles).  
(LO in a perturbative expansion vanishes for these channels.)
The empirical phase shifts from the SAID program~\cite{SAID} (solid circles)
are shown for comparison.
Reprinted figure with permission from Ref. \cite{Wu:2018lai}.
Copyright (2019) by the American Physical Society.}
\label{fig:3P2Bands}
\end{figure}

The main phenomenological shortcomings of the renormalized approach 
are $^3P_1$, $^3P_2$ and singlet partial waves.
For most of these channels, subsequent work indicates OPE might
be perturbative. 
Equation \eqref{MNNs=0} shows that OPE should be included
in $^1P_1$ at NLO, in $^1D_2$ at N$^2$LO, and so on.
On the basis of NDA, contact interactions with the minimal number 
of derivatives are expected at respectively N$^2$LO, N$^4$LO, and so on.
Under the assumption that the angular-momentum suppression
of TPE is the same as OPE,
Ref. \cite{Wu:2018lai} provided evidence that 
the perturbative expansion converges for singlet waves up to
$k\approx 300$ MeV and N$^4$LO without explicit Delta isobars.
Reference \cite{Wu:2018lai} goes further by showing
that under NDA for the LECs
also triplet waves converge in the same range,
except for $^3P_0$ and possibly $^3D_3$.
For illustration, results for the $^3P_2$-$^3F_2$ coupled channels
are shown in Fig. \ref{fig:3P2Bands} \cite{Wu:2018lai}, which should be compared
to Fig. \ref{renormChEFTLO} where OPE was treated nonperturbatively at LO.
The maximum momentum $k\approx 300$ MeV
seems tied to the absence of an explicit Delta isobar
\cite{Wu:2018lai} but no similar calculation is available in Deltaful ChEFT.
Earlier studies \cite{Kaiser:1997mw,Ballot:1997ht,Kaiser:1998wa},
which indicated that pions are perturbative in high waves,
sometimes included Deltas but did not take into account
the IR enhancement in iterated pion exchange.
Clearly a more comprehensive study of higher orders with
Deltas is needed
to confront this renormalized approach with phenomenology.

The situation is particularly unsatisfactory in the $^1S_0$ channel,
where LO --- same as in Weinberg's prescription at fixed pion mass ---
is far off, just as in Pionless EFT \cite{Hammer:2019poc}.
In particular, the Nijmegen PWA displays a zero at a relative low
momentum $k_0\simeq 340$ MeV, which is absent at LO. 
It is possible that the inclusion of an explicit Delta isobar
(separated in mass from the nucleon by $\sim 300$ MeV)
improves the convergence in this region, as a large
part of the central potential moves from N$^3$LO to N$^2$LO.
However, the expansion will in any case converge at best very slowly
for $k\simge k_0$, as all subleading orders have to conspire
to cancel against LO. 
Since numerically $k_0\sim M_{N\!N}$, only for a fully
perturbative-pion approach is this of no concern.
Note that also $^3S_1$ and $^3P_0$ have amplitude zeros at relatively
low energies, but in both cases they arise at LO
from the combination of nonperturbative
OPE and contact interactions need for renormalization.

\begin{figure}[t]
\begin{center}
\includegraphics[scale=.4]{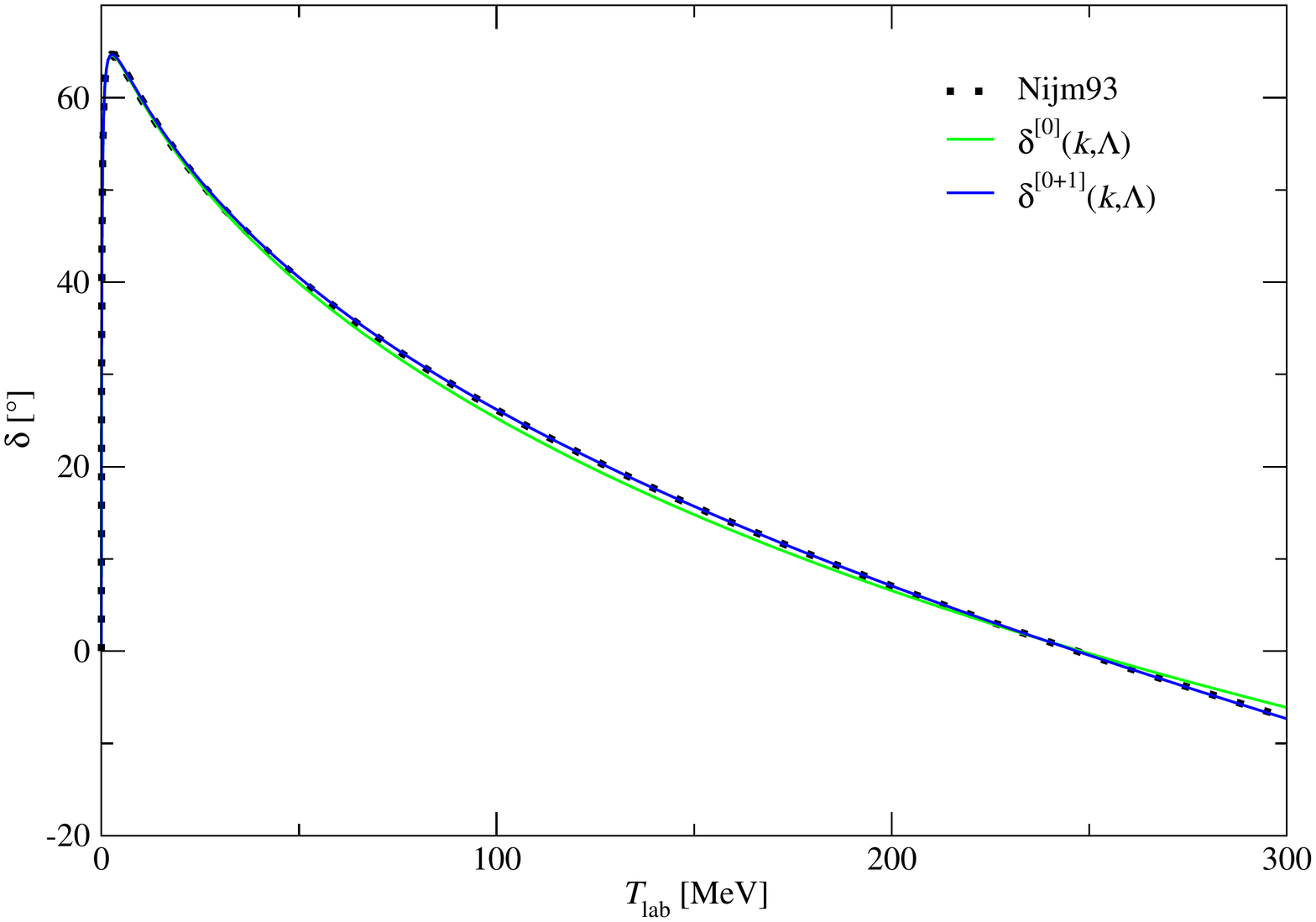}
\end{center}
\caption{Two-nucleon $^1S_0$ phase shift $\delta$ as function
of the laboratory energy $T_{\rm lab}$ in an expansion that incorporates 
the amplitude zero at LO.
The LO (green) and NLO (blue) bands correspond to cutoff variation
from 0.6 to 2 GeV. 
The results from the Nijm93 potential \cite{Stoks:1994wp} (black squares) are
shown for comparison.
Reprinted figure with permission from Ref. \cite{SanchezSanchez:2017tws}.
Copyright (2018) by the American Physical Society.}
\label{1S0withzero}
\end{figure}

The $^1S_0$ channel is special also for the presence of an unnaturally
shallow virtual state that requires a fine-tuning of the short-range
interaction. It is the interference between the
non-derivative contact interaction
and the Yukawa potential that causes a violation of NDA in this channel.
It also leads to the piling up of higher-order counterterms seen in 
Table \ref{contactorders}.
Given the uniqueness of this channel, it is perhaps not surprising that 
power counting might require refinement. In Ref. \cite{Birse:2010jr}
it was shown that short-range 
interactions show strong energy dependence. 
To ameliorate the expansion in $^1S_0$, it
was suggested in Refs. \cite{Beane:2001bc,Long:2013cya}
that the chirally symmetric two-derivative
interaction with LEC $C_{2(0)}$ should be promoted from NLO to LO,
following an earlier suggestions for Pionless EFT \cite{Beane:2000fi}
and ChEFT with purely perturbative pions~\cite{Ando:2011aa}.
To avoid the Wigner bound,
this is done through a dibaryon field \cite{Kaplan:1996nv}
whose kinetic term is taken to be LO
together with its residual mass. 
This promotion induces promotions at higher orders 
of the contact interactions with more derivatives.
Results of course improve at LO and further at NLO, but not
at N$^2$LO, in particular near $k_0$.
In Ref. \cite{SanchezSanchez:2017tws} it was then proposed
--- similarly to an earlier attempt \cite{Lutz:1999yr} ---
that the zero be included at LO by a combination of dibaryon field
and contact interaction (or alternatively two dibaryon fields,
the kinetic term of one of which is higher order).
Again this induces the promotion of contact interactions
with more derivatives at higher orders.
Phase shifts come out great at LO and essentially on the nose
at NLO, even beyond $k_0$, see Fig. \ref{1S0withzero}.
Unfortunately these reorganizations of the expansion 
produce energy-dependent potentials at LO, which complicate
few-body calculations.

A further proposed reorganization of ChEFT arises from treating 
selected relativistic corrections, which are small for the momenta of interest,
as LO --- see, for example, Ref. \cite{Behrendt:2016nql}.
A modified nucleon propagator ensures less dependence on the regulator,
but a $^3P_0$ LEC still has to be promoted compared to NDA, as in the purely
nonrelativistic context~\cite{Nogga:2005hy}.
By resumming higher-order terms into LO
whether they are relativistic corrections or not, one can soften
the large-momentum behavior of loops and alter the cutoff dependence.
This is no different than picking a regulator, which effectively includes
an infinite number of higher-derivative interactions.
Results then depend on the corresponding cutoff parameter $\Lambda$.
Renormalization exchanges this dependence for the minimal number of 
parameters allowed without dynamical assumptions.
Achieving cutoff independence with a resummation of a selected interaction
merely replaces $\Lambda$ by the mass parameter characterizing 
this interaction, call it $M'$.
If $M'\ll \MQCD$ is inferred from data,
this resummation is justified because the interaction is not 
of higher order.
However, when resumming relativistic corrections $M'\simge m_N$:
it corresponds to one fixed cutoff value and 
convergence cannot be used to demote interactions that
are needed for renormalization without resummation.
As long as no LECs are promoted or demoted,
a resummation of higher-order corrections is safe.
There is growing interest in the development of 
a covariant version of ChEFT, which could perhaps be used as input
to relativistic formulations of nuclear physics
\cite{Ren:2016jna,Ren:2017yvw}.

\subsection{More Nucleons}
\label{chiEFTmoreN}

There is not much known about renormalized ChEFT beyond $2N$.
The power counting of Sec. \ref{lessons} shows that the $3N$ force
is expected to start at NLO from two-pion exchange 
when Delta isobars are included explicitly,
and at N$^2$LO when they are not.
The crucial issue is whether shorter-range interactions are enhanced as in 
the $2N$ system. 
Such an enhancement does take place in Pionless EFT \cite{Bedaque:1999ve}
and it has been suggested for ChEFT on phenomenological grounds in 
Ref. \cite{Kievsky:2016kzb}.

In calculations for more than two nucleons in the renormalized approach,
one needs to truncate the LO $2N$ potential for $l \simle l_{\rm cr}^{(s)}$,
which is reminiscent of the truncation in total $2N$
angular momentum typically 
invoked in solutions of the Faddeev and Faddeev-Yakubovski equations for 
$3N$ and $4N$ systems with phenomenological potentials.
As we have seen the optimal values for $l_{\rm cr}^{(0,1)}$ are uncertain 
and the $l$ dependence of $M_{N\!N}^{(l,s)}$ is not fully determined.
Of course, as in the $2N$ system, subleading orders should
be treated in distorted-wave perturbation theory.

\begin{figure}[t]
\begin{center}
\includegraphics[scale=.4]{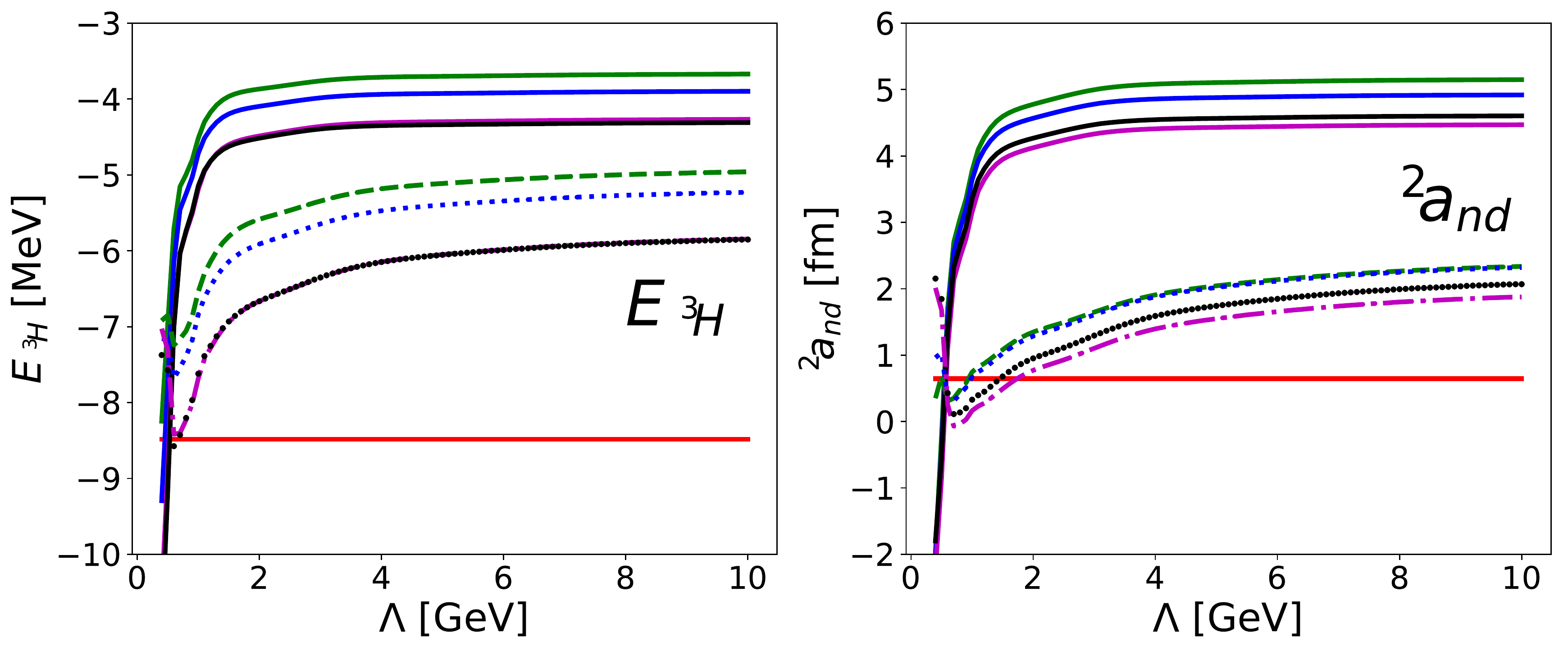}
\end{center}
\caption{Triton binding energy $E_{^3\! H}$ and 
doublet neutron-deuteron scattering length $^2a_{nd}$ 
as functions of the cutoff $\Lambda$.
Results at LO (solid lines) and NLO (dashed and dotted lines) 
for various $2N$ fitting procedures are compared with
experiment (horizontal red lines).
Reprinted figure with permission from Ref. \cite{Song:2016ale}.
Copyright (2019) by the American Physical Society.}
\label{3N}
\end{figure}

Existing calculations are limited to the $3N$ system and took 
$l_{\rm cr}^{(0,1)}=3$. 
At LO~\cite{Nogga:2005hy,Song:2016ale} and, without 
explicit Deltas, also at NLO~\cite{Song:2016ale},
observables converge as the cutoff increases to at least 10 GeV
without $3N$ forces, see Fig. \ref{3N}~\cite{Song:2016ale}.
The triton binding energy is 
$B_3^{\rm LO}\simeq 4$ MeV and $B_3^{\rm NLO}\simeq 6$ MeV,
quite different from results for a low cutoff in 
Weinberg's prescription, $\simeq 11$ MeV ($\simeq 6.5$ MeV)
at LO (N$^2$LO) \cite{Lynn:2017fxg}.
Results were shown not to change significantly 
when waves beyond $l_{\rm cr}=3$ were included.
Conversely, if it turns out that $l_{\rm cr}^{(0,1)}<3$, results might change 
quantitatively, but qualitative statements should stand.
In particular, one concludes there is no renormalization justification
in ChEFT to take the non-derivative $3N$ contact interaction as LO.
Most likely the same conclusion holds for higher-body forces,
but no calculations have been carried out.

The tendency for underbinding at LO seen in the deuteron and triton 
seems to persist for symmetric nuclear matter.
In a cutoff-converged Brueckner pair approximation
\cite{Machleidt:2009bh}, nuclear matter was found to saturate,
but with significant underbinding. 
This is in contrast to Weinberg's prescription,
where Deltaless \cite{Sammarruca:2018bqh} or Deltaful \cite{Ekstrom:2017koy}
potentials of ${\cal O}(1)$ and ${\cal O}(Q^2/\MQCD^2)$
do not yield saturation within the EFT domain. 
Yet higher potentials do lead to saturation with this prescription 
\cite{Ekstrom:2017koy,Drischler:2017wtt,Sammarruca:2018bqh}.
Although usually presented as a success,
the emperor has no clothes: it means that,
if nuclear matter is within the regime of ChEFT,
interactions that are formally
of higher order according to NDA must actually be LO to balance
against other LO interactions.
Presumably it is the extra repulsion from $^3P_0$ in a renormalized approach
that saturates nuclear matter.
It is not clear how saturation in Chiral EFT would relate, if it can be
related at all, to the proposal of Ref. \cite{Kolck:2017zzf}
where saturation arises from the $3N$ parameter that
appears at LO in Pionless EFT.
What is clear is,
more EFT calculations beyond the $2N$ system are sorely needed!

\section{Conclusion}
\label{conc}

The longstanding problem of renormalization of chiral nuclear forces
has been solved at the $2N$ and $3N$ levels.
Perhaps not surprising in hindsight, this solution is a middle 
ground between Weinberg's original prescription and 
Kaplan, Savage and Wise's suggestion of fully perturbative pions. 
One-pion exchange is iterated in lower waves together with the necessary
contact interactions, while all corrections are included in
distorted-wave perturbation theory.

That is not to say that the {\it best} solution has been found.
Issues remain regarding exactly how strong the angular-momentum suppression is
and where the nonperturbative/perturbative boundary lies.
Whether the ordering of few-body forces holds similar surprises is also unknown.
A high-quality fit to $2N$ data is missing,
and there are very few studies of heavier systems.
The extent to which Weinberg's phenomenologically successful
prescription with a low cutoff can be reproduced remains an open question,
although the first step in grounding it on a renormalized approach
has been made \cite{Valderrama:2019lhj}.
Fortunately, there is still plenty to learn.

\section*{Acknowledgments}
I thank Manolo Pav\'on Valderrama for useful discussions
and Jaber Balal Habashi for comments on the manuscript.
This material is based upon work supported in part 
by the U.S. Department of Energy, Office of Science, Office of Nuclear Physics, 
under award DE-FG02-04ER41338
and by the European Union Research and Innovation program Horizon 2020
under grant No. 654002.

\end{document}